\documentclass[journal]{IEEEtran}

\IEEEoverridecommandlockouts

\usepackage{algorithmic}
\usepackage{graphicx}
\usepackage{textcomp}
\usepackage{xcolor}
\usepackage{siunitx}
\usepackage{float}
\usepackage{stfloats}
\usepackage{tabularx}
\usepackage{textgreek}
\usepackage{hhline}
\usepackage{mathtools}
\usepackage{booktabs}
\usepackage[percent]{overpic}
\usepackage{listings}
\usepackage{wrapfig}
\usepackage{tikz}
\usepackage{pgfplots}
\DeclareUnicodeCharacter{2212}{−}
\usepgfplotslibrary{groupplots,dateplot}
\usetikzlibrary{patterns,shapes.arrows}
\pgfplotsset{compat=newest}

\usetikzlibrary{shapes,arrows}
\usetikzlibrary{circuits.logic.US} 

\makeatletter
\let\old@lstKV@SwitchCases\lstKV@SwitchCases
\def\lstKV@SwitchCases#1#2#3{}
\makeatother
\usepackage{lstlinebgrd}
\makeatletter
\let\lstKV@SwitchCases\old@lstKV@SwitchCases

\lst@Key{numbers}{none}{%
    \def\lst@PlaceNumber{\lst@linebgrd}%
    \lstKV@SwitchCases{#1}%
    {none:\\%
     left:\def\lst@PlaceNumber{\llap{\normalfont
                \lst@numberstyle{\thelstnumber}\kern\lst@numbersep}\lst@linebgrd}\\%
     right:\def\lst@PlaceNumber{\rlap{\normalfont
                \kern\linewidth \kern\lst@numbersep
                \lst@numberstyle{\thelstnumber}}\lst@linebgrd}%
    }{\PackageError{Listings}{Numbers #1 unknown}\@ehc}}
\makeatother

\definecolor{highlight}{RGB}{233,172,169} 

\usepackage{fancyhdr}

\usepackage[utf8]{inputenc}

\newcolumntype{Y}{>{\centering\arraybackslash}X}

\usepackage[acronym]{glossaries}

\definecolor{codegreen}{rgb}{0,0.6,0}
\definecolor{codegray}{rgb}{0.5,0.5,0.5}
\definecolor{codepurple}{rgb}{0.58,0,0.82}
\definecolor{backcolour}{rgb}{0.95,0.95,0.92}

\lstdefinestyle{mystyle}{
    commentstyle=\color{codegreen},
    keywordstyle=\color{magenta},
    numberstyle=\tiny\color{codegray},
    stringstyle=\color{codepurple},
    basicstyle=\ttfamily\footnotesize,
    breakatwhitespace=false,         
    breaklines=true,                 
    captionpos=b,                    
    keepspaces=true,                 
    numbers=left,                    
    numbersep=5pt,                  
    showspaces=false,                
    showstringspaces=false,
    showtabs=false,                  
    tabsize=1
}

\newacronym[plural=CNNs, firstplural={Convolutional Neural Networks (CNNs)}]{cnn}{CNN}{Convolutional Neural Network}
\newacronym[plural=BNNs, firstplural={Binary Neural Networks (BNNs)}]{bnn}{BNN}{Binary Neural Network}
\newacronym[plural=TNNs, firstplural={Ternary Neural Networks (TNNs)}]{tnn}{TNN}{Ternary Neural Network}
\newacronym[plural=NNs, firstplural={Neural Networks}]{nn}{NN}{Neural Network (NNs)}
\newacronym[plural=SCMs, firstplural={Standard Cell Memories (SCMs)}]{scm}{SCM}{Standard Cell Memory}
\newacronym{ann}{ANN}{Artificial Neural Networks}
\newacronym{ml}{ML}{Machine Learning}
\newacronym{iot}{IoT}{Internet of Things}
\newacronym{fft}{FFT}{Fast Fourier Transform}
\newacronym[plural=OCUs, firstplural={Output Channel Compute Units (OCUs)}]{ocu}{OCU}{Output Channel Compute Unit}
\newacronym{alu}{ALU}{Arithmetic Logic Unit}
\newacronym{fifo}{FIFO}{First-In First-Out Queue}
\newacronym{asic}{ASIC}{Application-Specific Integrated Circuits}
\newacronym{mac}{MAC}{Multiply-Accumulate}
\newacronym{vcd}{VCD}{Value Change Dump}
\newacronym{soc}{SoC}{System-on-Chip}

\definecolor{darkred}{HTML}{A8322C}
\newcommand{\revA}[1]{\textcolor{black}{#1}}
\lstdefinestyle{myPython}{
    keywordstyle=\bfseries\color{black},
    basicstyle=\ttfamily\footnotesize,
    breakatwhitespace=false,         
    breaklines=true,                 
    captionpos=b,                    
    keepspaces=true,                 
    numbers=left,                    
    numbersep=5pt,                  
    showspaces=false,                
    showstringspaces=false,
    showtabs=false,                  
    tabsize=2
}

\begin{document}
\title{CUTIE: Beyond PetaOp/s/W Ternary DNN Inference Acceleration with Better-than-Binary Energy Efficiency}

\author{Moritz Scherer, Georg Rutishauser, Lukas Cavigelli, Luca Benini
\thanks{M. Scherer, G. Rutishauser, and L. Benini are with the Dept. of Information Technology and Electrical Engineering, ETH Z\"{u}rich, Switzerland (e-mail: 
\{scheremo, georgr, benini\}@iis.ee.ethz.ch).} 
\thanks{L. Cavigelli is with Huawei Technologies, Zurich Research Center, Switzerland (e-mail: lukas.cavigelli@huawei.com).}
\thanks{L. Benini is also with the Dept. of Electrical, Electronic and Information Engineering, University of Bologna, Italy.}
}

\markboth{
Under Review at IEEE Transactions on Computer-Aided Design of Integrated Circuits and Systems}{
M. Scherer \MakeLowercase{\textit{et al.}}: CUTIE: Beyond PetaOp/s/W Ternary DNN Inference Acceleration with Better-than-Binary Energy Efficiency}

\maketitle

\begin{abstract}
We present a 3.1 POp/s/W fully digital hardware accelerator for ternary neural networks. CUTIE, the Completely Unrolled Ternary Inference Engine, focuses on minimizing non-computational energy and switching activity so that dynamic power spent on storing (locally or globally) intermediate results is minimized. This is achieved by 1) a data path architecture completely unrolled in the feature map and filter dimensions to reduce switching activity by favoring silencing over iterative computation and maximizing data re-use, 2) targeting ternary neural networks which, in contrast to binary NNs, allow for sparse weights which reduce switching activity, and 3) introducing an optimized training method for higher sparsity of the filter weights, resulting in a further reduction of the switching activity.
Compared with state-of-the-art accelerators, CUTIE achieves greater or equal accuracy while decreasing the overall core inference energy cost by a factor of 4.8$\times$--21$\times$.

\end{abstract}



\section{Introduction}

Since the breakthrough success of AlexNet in the ILSVRC image recognition challenge in 2012 \cite{Krizhevsky2012}, \glspl{cnn} have become the standard algorithms for many machine learning applications, especially in the fields of audio and image processing. Supported by advances in both hardware technology and neural network architectures, dedicated \gls{asic} hardware accelerators for inference have become increasingly commonplace, both in datacenter-scale applications as well as in consumer devices \cite{Reuther2019}. 
With the increasing demand to bring machine learning to \gls{iot} devices and sensor nodes at the very edge, the \textit{de facto} default paradigm of cloud computing is being challenged. Neither are most data centers able to process the sheer amount of data generated by billions of sensor nodes nor can typical edge devices afford to send their raw sensor data to data centers for further processing, given their very limited power budget \cite{Chen2019}. 
One solution to this dilemma is to increase the processing capabilities of each sensor node to enable it to only send extracted, highly compressed information over power-intensive wireless communication interfaces or to act as an autonomous system.

However, the general-purpose microcontrollers typically employed in these \gls{iot} devices are ill-suited to the computationally intensive task of DNN inference, placing severe limitations on the achievable energy efficiency. While great strides in terms of energy efficiency have been made with specialized microcontrollers \cite{Gautschi2017}, some applications still require lower power consumption than what can be achieved with using 32-bit weights and activations in DNN inference. 
A popular approach to reducing the power consumption for neural network computations is the quantization of network parameters (weights) and intermediate results (activations). 
Quantized inference at a bit-width of 8 bits has been shown to offer equivalent statistical accuracy while allowing for significant savings in computation energy as well as reducing the requirements for working memory space, memory bandwidth, and storage by a factor of 4 compared to traditional 32-bit data formats \cite{Han2015deep, Han2015learning, Iandola2016, Cavigelli2020EBPC}. 

Pushing along the reduced bit-width direction, recently several methods to train neural networks with binary and ternary weights and activations have been proposed \cite{Courbariaux2015, Rastegari2016, Hubara2016, Li2016, Zhu2016, Lin2017}, allowing for an even more significant decrease in the amount of memory required to run inference. In the context of neural networks, binary values refer to the set \{-1, +1\} and ternary values refer to the set \{-1, 0, 1\} \cite{Courbariaux2015, Alemdar2017}. 
These methods have also been used to convert complex state-of-the-art models to their Binary Neural Network (BNN) or Ternary Neural Network (TNN) form. While this extreme quantization incurs sizeable losses in accuracy compared to the full-precision baselines, such networks have been shown to work well enough for many applications and the accuracy gap has been reducing quite rapidly over time \cite{Qin2020, Mishra2017, Choi2018}. 

Although quantization of networks does not affect the total number of operations for inference, it reduces the complexity of the required multipliers and adders, which leads to much lower energy consumption per operation. For binary networks, a multiplier can be implemented by a single XNOR-gate \cite{Andri2020}. Further, the number of bit accesses per loaded value is minimized, which not only reduces the memory footprint but also the required wiring and memory access energy.

While \glspl{bnn} in particular are fairly well-suited to run on modern general-purpose computing platforms, to take full advantage of the potential energy savings enabled by aggressively quantized, specialized, digital, low-power hardware accelerators have been developed \cite{Andri2016, Andri2020, Moons2018, Chen2018}.  
Concurrently to the research in digital neural network accelerators, analog accelerators that compute in-memory, as well as mixed-signal, have been explored \cite{Jain2019, Valavi2019, Bankman2019}. While mixed-signal and in-memory designs hold the promise of higher energy efficiency than purely digital designs under nominal conditions, their higher sensitivity to process and noise variations, coupled with the necessity of interfacing with the digital world, are open challenges to achieve their full potential in energy efficiency \cite{Klachko2019}.

Even though both analog and digital accelerators extract immense performance gains from the reduced complexity of each operation, there is still untapped potential to further increase efficiency.
Most state-of-the-art binary accelerators use arrays of multipliers with large adder trees to perform the multiply-and-popcount operation \cite{Andri2020, Moons2018, Knag2020, Bankman2019}, which induces a large amount of switching activity in the adder tree, even when only a single input node is toggled. 
Adding to this, even state-of-the-art binary accelerators spend between 30\% to 70\% of their energy budget on data transfers from memories to compute units and vice-versa \cite{Bankman2019, Moons2019}. This hurts efficiency considerably since time and energy spent on moving data from memories to compute units are not used to compute results.
Taking these considerations into account, two major opportunities for optimization are to reduce switching activity in the compute units, especially the adder trees, and to reduce the amount of data transfer energy.

In this paper, we explore three key ideas to increase the core efficiency of digital low-bit-width neural network accelerator architectures: first, \textit{unrolling of the data-path architecture with respect to the feature map and filter dimensions} leading to lower data transfer overheads and reduced switching activity compared to designs that implement iterative computations. Second, \textit{focusing on \glspl{tnn} instead of \glspl{bnn} thereby capitalizing on sparsity to statistically decrease switching activity in unrolled compute units}. Third, \textit{optimizing the quantization strategy of \glspl{tnn} resulting in sparser networks that can be leveraged with an unrolled architecture}.
We combine these ideas in CUTIE, the \textit{Completely Unrolled Ternary Inference Engine}.

Our contributions to the growing field of energy-optimized aggressively quantized neural network accelerators are as follows:
\begin{enumerate}
    \item We present the design and implementation of a novel accelerator architecture, which minimizes data movement energy spending by unrolling the compute architecture in the feature map and filter dimensions, demonstrating that non-computational energy spending can be reduced to less than 10\% of the overall energy budget (Section \ref{sec:experimentalresults}).
    \item We demonstrate that by unrolling each compute unit completely and adjusting the quantization strategy, we directly exploit sparsity, minimizing switching activity in multipliers and adders, reducing the inference energy cost of ternarized networks by 36\% with respect to their binarized variants (Section \ref{sec:comparisonquantization}).
    \item We present analysis results, showing that the proposed architecture achieves up to \SI{589} {TOp/s/W} in an IoT-suitable \SI{22}{\nano\meter} technology and up to \SI{3.1}{POp/s/W} in an advanced \SI{7}{\nano\meter} technology, outperforming the state-of-the-art in digital, as well as analog in-memory BNN accelerators, by a factor of $4.8\times$ in terms of energy per inference at iso-accuracy (Section \ref{sec:comparisonsoa}).
\end{enumerate}
This paper is organized as follows: in Section \ref{sec:relatedwork}, previous work in the field of neural network hardware accelerators and aggressively quantized neural networks is discussed. In Section \ref{sec:architecture}, we introduce the proposed accelerator architecture. Section \ref{sec:implementation} details the implementation of the architecture in the GlobalFoundries \SI{22}{\nano\meter} FDX and TSMC \SI{7}{\nano\meter} FF technologies. In Section \ref{sec:results}, the implementation results are presented and discussed, by comparing with previously published accelerators. Finally, Section \ref{sec:conclusion} concludes this paper, summarizing the results.

\section{Related Work}\label{sec:relatedwork}


In the past few years, considerable research effort has been devoted to developing task-specific hardware architectures that enable both faster neural network inference as well as a reduction in energy per inference.
A wide range of approaches to increase the energy-efficiency of accelerators have been studied, from architectural and device-level optimizations to sophisticated co-optimization of the neural network and the hardware platform. 

\subsection{Aggressively Quantized Neural Networks}

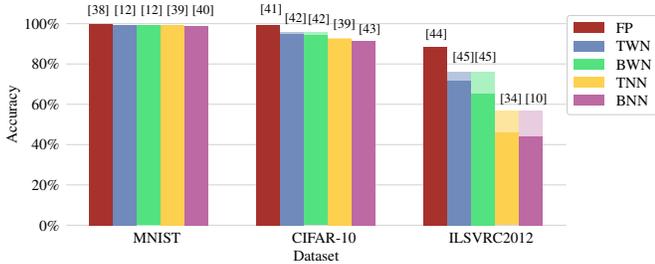
\begin{figure}
    \begin{center}
    \begin{overpic}[width=\linewidth,percent,trim=0 20 0 0]{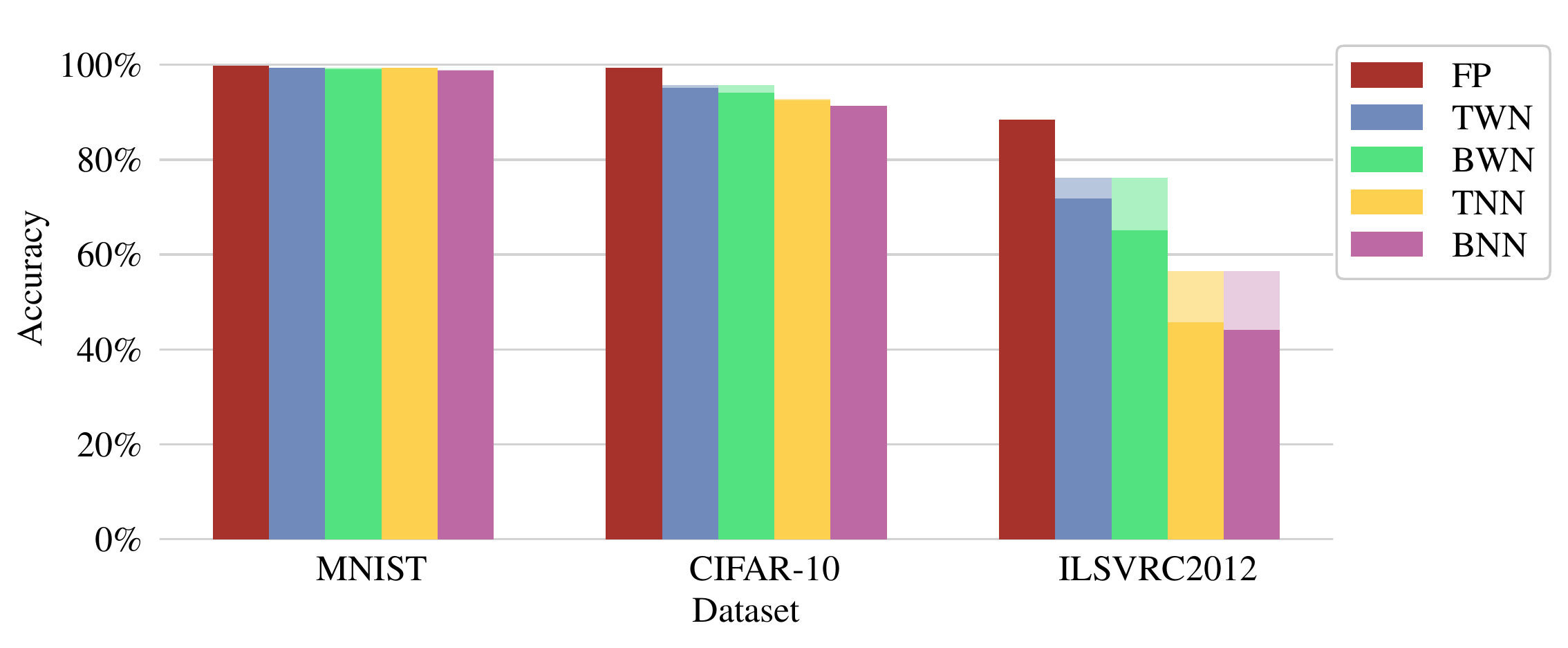}
        \put(13.3, 35.5){\tiny \cite{ref:branch_merge}}
        \put(17.3, 35.5){\tiny \cite{Li2016}}
        \put(21.3, 35.5){\tiny \cite{Li2016}}
        \put(25.2, 35.5){\tiny \cite{ref:gxnor-net}}
        \put(29.0, 35.5){\tiny \cite{ref:xnor-rram}}
        \put(39.2, 35.7){\tiny \cite{ref:bigtransfer}}
        \put(42.9, 34.4){\tiny \cite{ref:binary_relax}}
        \put(46.4, 34.25){\tiny\cite{ref:binary_relax}}
        \put(50.2, 33.5){\tiny \cite{ref:gxnor-net}}
        \put(54.0, 32.75){\tiny \cite{ref:bnn_plus}}
        \put(64.0, 32.0){\tiny \cite{ref:fix_eff_net}}
        \put(68.1, 28.5){\tiny \cite{ref:rpr}}
        \put(71.4, 28.5){\tiny \cite{ref:rpr}}
        \put(75.2, 22.3){\tiny \cite{ref:ana}}
        \put(79.0, 22.3){\tiny \cite{Rastegari2016}}
        \end{overpic}
    \end{center}
    \caption{\revA{Comparison of state-of-the-art accuracy of highly quantized neural networks of different precisions. \textbf{FP}: state-of-the-art in unquantized/full-precision neural networks, \textbf{BWN/TWN}: binary/ternary weight networks,  \textbf{BNN/TNN}: fully binarized/ternarized neural networks. For the quantized network categories, the accuracy of the corresponding unquantized baseline networks is shown greyed out. As task difficulty is increased, a) the performance gap between the quantized networks and the full-precision baselines increases, and b) the gap between the unquantized baselines from which the quantized architectures are derived and the full-precision state-of-the-art widens.}}
    \label{fig:quantized_acc}
\end{figure}

On the algorithmic side, one of the main recent research directions has been quantization, i.e. representing model weights and intermediate activations in lower arithmetic precision. It has been known for some time that quantization of network weights to \SI{5}{bits} and less is possible without a loss in accuracy in comparison to a 32-bit floating-point baseline model \cite{Han2015deep, Han2015learning, Iandola2016}. Further quantization of network weights to binary or ternary precision usually results in a small drop in accuracy, but precision is still adequate for many applications \cite{Li2016, Zhu2016, Hu2018, cerutti2020sound}. 
Extending the approach of extreme quantization to intermediate activations, fully binarized and fully ternarized networks have been proposed \cite{Courbariaux2015, Alemdar2017}.  These types of networks perform very well on easier tasks such as 10-class classification on the well-established MNIST dataset \cite{Deng2017}, and efforts have been taken to improve their performance with novel training approaches \cite{Zhou2017, Bulat2019, ref:ana}. Nevertheless, on more challenging tasks such as classification on the ILSVRC’12 dataset, they are still significantly less accurate than their full-precision counterparts \cite{Rastegari2016, Zhou2016, Hubara2016, Mishra2017, Lin2017, Zhuang2019, Phan2020}. 
Figure \ref{fig:quantized_acc} depicts the accuracy gap between previously published, strongly quantized neural networks, their full-precision equivalents with identical architectures and the state-of-the-art full-precision networks on  image classification tasks of increasing difficulty. On higher difficulty tasks, the gap between quantized networks and their full-precision equivalents grows larger. Furthermore, the gap between the full-precision architectures from which the quantized networks are derived and the overall state-of-the-art results reported in literature grows with task difficulty, indicating a prevalent focus in research activity on easier tasks and simple networks.


Taking all of this into account, \glspl{bnn} and \glspl{tnn} provide a unique and interesting operating point for embedded devices, since they are by definition aggressively compressed, allowing for deep model architectures to be deployed to highly memory-constrained low-power embedded devices. 

The core idea of binarization and ternarization of neural networks has been applied in numerous efforts, some of which also study the impact of the quantization strategy on the sparsity of ternary weight networks \revA{\cite{Zhu2016, Faraone2017, Marban2020, Ding2019}}. While these previous efforts focus on the impact of the choice of quantization threshold and regularization, we evaluate the impact of quantization order, rather than threshold or regularization. Further, we study the effect of sparsity on the energy-efficiency of the proposed accelerator architecture.














\subsection{DNN Hardware Accelerators}



While the first hardware accelerators used for neural networks were general-purpose GPUs, there has been a steady trend pointing towards specialized hardware acceleration in machine learning in the past few years \revA{\cite{Sze2017, 10.1145/2744769.2744788, Chen2020, Fowers2018}}. 
Substantial research efforts have focused on exploring efficient architectures for networks using activations and weights with byte-precision or greater, \cite{Chen2016, Jouppi2017, Moons2017, Chen2018} different digital \gls{asic} implementations for binary weight networks and \gls{bnn}s have been proposed \cite{Andri2016, Moons2018, Andri2018, Conti2019, ref:ebpc_conf, Andri2020}. Some works have tackled analog \gls{asic} implementations of \gls{tnn} accelerators, \cite{Jain2019, Okumura2019}, but very few digital implementations for \gls{tnn} accelerators have been published \cite{Ando2018, Ardakani2018}. 

At the heart of every digital neural network accelerator lie the processing elements, which typically compute \gls{mac} operations. An important distinction between different architectures, besides the supported precision of their processing elements, lies in the way they schedule computations \cite{Sze2017}. 
Most state-of-the-art architectures can be categorized into systolic arrays \cite{Chen2016, Conti2018, Andri2018, Chen2018, Jain2019}, which are flexible in how their processing elements are used, or output-stationary designs, which assign each output channel to one processing element \cite{Sze2017, Moons2018, Knag2020}. Both approaches trade-off lower area for lower throughput and increased data transfer energy by using iterative decomposition since partial results need to be stored and either weights or feature map data need to be reloaded. The alternative to iterative decomposition pursued in our approach, i.e. fully parallelizing the kernel-activation dot-products, is not only generally possible for convolutional neural networks, but also promises to be more efficient by increasing data-reuse and parallelism. 

The state-of-the-art performance in terms of energy per operation for digital \gls{bnn} and \gls{tnn} accelerators is reported in Moons et al. \cite{Moons2018} and Andri et al. \cite{Andri2020}, achieving peak efficiencies of around \SI{230}{TOp/s/W} for 1-bit operations, as well as Knag et al. \cite{Knag2020}, reporting up to \SI{617}{TOp/s/W}. The state-of-the-art for ternary neural networks is found in Jain et al. \cite{Jain2019}, achieving around \SI{130}{TOp/s/W} for ternary operations.

In this work, we move beyond the state-of-the-art in highly quantized acceleration engines by implementing a completely unrolled data path. We show that by unrolling the data path, sparsity in \glspl{tnn} is naturally exploited to reduce the required energy per operation without any additional overhead, unlike previous works \cite{Moons2016, Sen2018, Zhou2019, Yuan2020}.
To capitalize on this effect, we introduce modifications to existing quantization strategies for \glspl{tnn}, which are able to extract 53\% more sparsity at iso-accuracy than by sparsity-unaware methods. Lastly, our work shows that ternary accelerators can significantly outperform binary accelerators both in terms of energy efficiency as well as statistical accuracy.






\section{System Architecture}\label{sec:architecture}

This section introduces the proposed system architecture. First, we present the data path and principle of operation and explain the levels of data re-use that the architecture enables, then we discuss considerations for lowering the overall power consumption. Finally, we present the supported functionality.

\subsection{High-level Data Path}

\begin{figure}
  \begin{center}    
    \includegraphics[width=\linewidth]{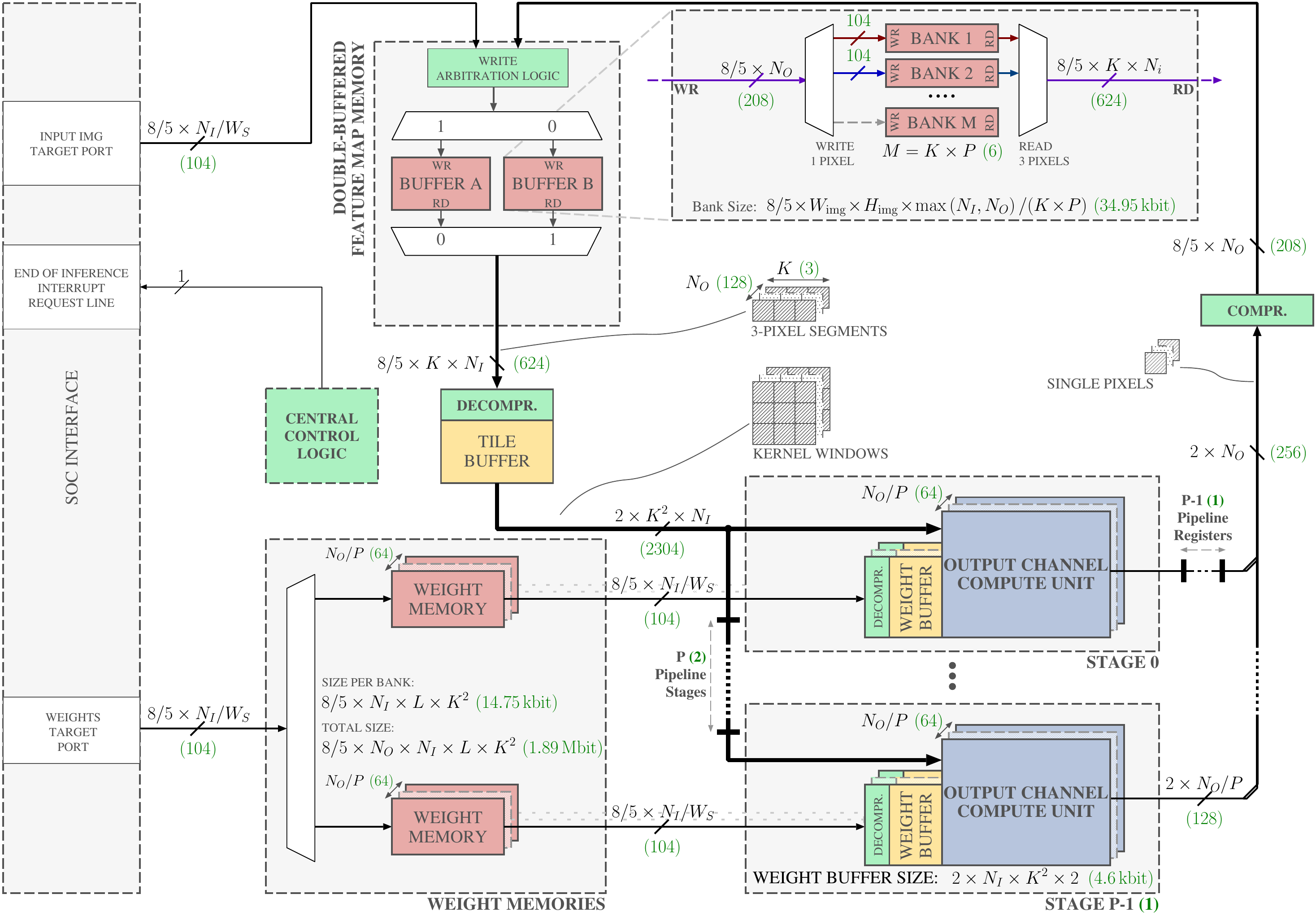}
    \caption{\revA{Data-path schematic view of the accelerator core and its embedding into an SoC-level system. The diagram shows the unrolled compute architecture and encoding/decoding blocks, as well as the weight and feature map memories and tile buffer module. The dataflow of the accelerator is scheduled to first buffer full feature map windows in the tilebuffer and then compute the convolution result with pre-loaded weights in the compute units after which they are saved back to the feature map memory.}
    }
    \label{fig:highlevelschematic}
  \end{center}
\end{figure}

Figure \ref{fig:highlevelschematic} shows a high-level block diagram of the accelerator architecture.
It is optimized for the energy-efficient layer-wise execution of neural networks. This is achieved first and foremost by a flat design hierarchy; each output feature map is computed channel-wise by dedicated compute units, called \gls{ocu}. Each \gls{ocu} is coupled with a private memory block for weight buffering, which minimizes addressing and multiplexing overheads for weight memory accesses, reducing the amount of energy spent on data transfers. The feature map storage buffers are shared between all \gls{ocu}s to maximize the re-use of loaded activation data, which again aims to decrease the data transfer energy. 

To exploit the high rate of data re-use possible with \glspl{cnn}, the design uses a tile buffer, which produces tiles, i.e. square windows, of the input feature map in a sliding window manner. These windows are then broadcast to the pipelined \glspl{ocu}. 

An important aspect of aggressively quantized and mixed-precision accelerator design is choosing a proper compression scheme for its values. Since ternary values encode $\mathrm{log}_{2}(3) \approx \SI{1.585}{bits}$ per symbol, the most straight-forward compression approach would require \SI{2}{bits} of memory per value, leaving one of the four possible codewords unused. To reduce this overhead, values are stored 5 at a time, using \SI{8}{bits} leading to \SI{1.6}{bits} per symbol. The compression scheme used for this representation is taken from a recent work by Muller et al. \cite{Muller2019}. To transition between the compressed representation and the standard 2's complement representation, compression and decompression banks are used with feature map and weight memories.

Figure \ref{fig:highlevelschematic} shows the pipeline arrangement of the \glspl{ocu}. A key feature of the architecture is that an output channel computation is entirely performed on a single \gls{ocu}. All \glspl{ocu} need to receive input activation layers: the broadcast of input activations to \glspl{ocu} is pipelined and the \glspl{ocu} are grouped in stages.  This pipeline fulfils multiple purposes: from a functional perspective, it allows to silence the input to clusters of compute units,  which reduces switching activity during the execution of layers with fewer output channels than the maximum. Concerning the physical implementation of the design, pipelining helps to reduce fanout, which further reduces the overall power consumption of the design. It also reduces the propagation delay introduced by physical delays due to long wires. 



\subsection{Parametrization}

The CUTIE architecture is parametrizable at compile time to support a large variety of design points. An overview of the design parameters is shown in Table \ref{tab:DesignParams}. Besides the parameters in Table \ref{tab:DesignParams}, the design's feature map memories and weight memories can be implemented using either \glspl{scm} or SRAMs. 
CUTIE is designed to support arbitrary odd square kernel sizes $K$, pipeline depths $P$, input channel numbers $N_{I}$ and output channel numbers $N_{O}$ which directly dictate the dimensioning of the compute core, but also of the feature map memories and the tile buffer.
The \gls{ocu}, as shown in Figure \ref{fig:ocu_pool}, consists of a compute core and a latch-based weight buffer that is designed to hold two kernels for the computation of one output channel, which amounts to $4 \times K^{2} \times N_{I} $ bits. 
The feature map memories are designed to support the concurrent loading of $K$ full pixels as well as the granular saving of $\frac{N_{O}}{P}$ ternary values. For these reasons, the word width of the feature map memories is chosen to be $\frac{N_{O}}{P}$ ternary values. To further allow for concurrent write and read accesses of up to $K$ pixels, two feature map memories, each with $P \times K$ feature map memory banks, are implemented. 

\begin{table}
  \caption[Design parameters of CUTIE]{Design parameters of CUTIE}
  \begin{tabularx}{\linewidth}{c|X}
    \textbf{Parameter} & \textbf{Description} \\
    \hline
    \revA{$N_I$} & Maximum number of channels of input feature map \\
    \revA{$N_O$} & Maximum number of channels of output feature map \\
    \revA{$K$} &  Maximum kernel width and height \\
    \revA{$I_W$} &  Maximum width of input feature map \\
    \revA{$I_H$} &  Maximum height of input feature map \\
    \revA{$L$} &  Maximum number of layers in the queue \\
    \revA{$P$} &  Number of pipeline stages \\
    \revA{$W_S$} & Number of memory words per pixel \\
  \end{tabularx}
  \label{tab:DesignParams}

\end{table}

\subsection{Principle of Operation}


\begin{figure}
        \centering
        \includegraphics[width=\linewidth, trim=0 20 0 0]{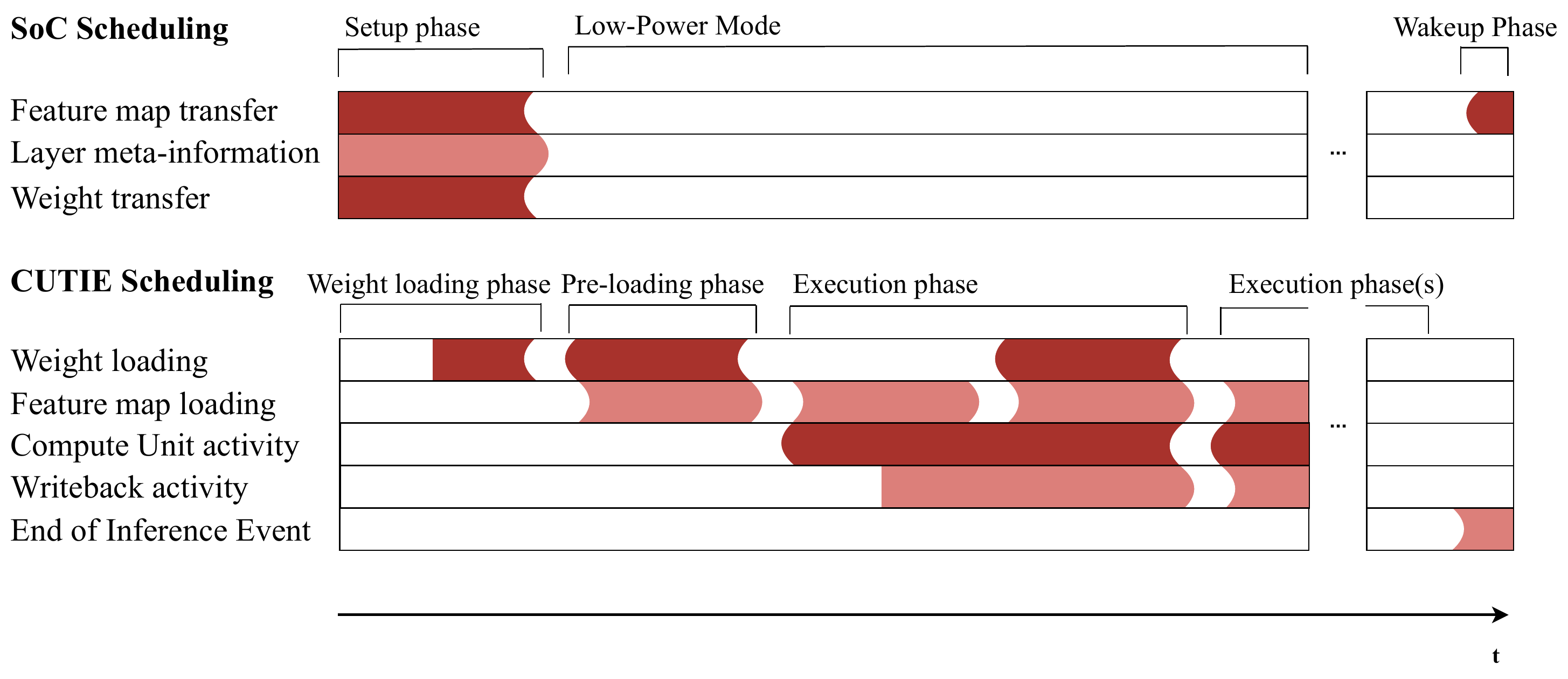}
        \caption{\revA{Scheduling diagram of the accelerator core and SoC interface. The first two phases are needed to set up the first layer after reset, every other loading phase overlaps with an execution phase, which reduces the latency for scheduling a new layer to a single cycle. The host system can be put in a low-power mode while the accelerator core computes the network since all layer information is saved inside the core's memories.}}
        \label{fig:schedulingdiagram}
\end{figure}

The accelerator core processes neural networks layer-wise. To enable layer-wise execution, networks have to be compiled and mapped to the core instruction set. The compilation process achieves two main goals: first, the networks' pooling layers are merged with the convolutional layers to produce fused convolutional layers. Second, the networks' convolutional layers' biases, batch normalization layers, and activation functions are combined to produce two thresholds that are used to ternarize intermediate results, similar to constant expression folding for \glspl{bnn} \cite{Conti2018}. After compilation, each layer consists of a convolutional layer with ternary weights, followed by optional pooling functions and finally, an activation function using two thresholds that ternarizes the result. To map the network to the accelerator, each layer's weights are stored consecutively in the weight memories, the thresholds are stored consecutively in the \gls{ocu}s' Threshold FIFO and the meta-information like input width, stride, kernel size, padding, and so on are stored in the layer FIFO. All FIFOs, controllers and scheduling modules combined make up 2\% of the total area.

The accelerator is designed to pre-buffer the weights for a full network during its setup phase and re-use the stored weights for multiple executions on different feature maps. 
Once at least one layer's meta-information is stored and the start signal is asserted, the accelerator's controllers schedule the execution of each layer in two phases; first, the weights for one layer are loaded into their respective buffers in the \glspl{ocu}, then the layer is executed, i.e. every sliding window's result is computed and written back to the feature map memory. The loading of weights into the \glspl{ocu} for the next layer and the computation of the current layer can overlap, leading to a single, fully concurrent execution phase after buffering the first set of weights, as shown in Figure \ref{fig:schedulingdiagram}. Once all layers have been executed, the end of inference signal is asserted, signalling to the host controller that the results are valid and the accelerator is ready for the next feature map input.

The module responsible for managing the loading and release of sliding windows is the tile buffer. 
The tile buffer consists of a memory array that stores $K$ lines of pixel values implemented with standard cell latches.
Feature maps are stored in a (H$\times$W$\times$C)-aligned fashion in the feature map memory. To avoid load stalls and efficiently feed data to the compute core, up to $K$ adjacent pixels at a time are read from the feature map memory.
The load address is computed to always target the leftmost pixel of a window.

The scheduling algorithm for the release of the windows keeps track of the central pixel of the next-to-be scheduled window. This can be used to enable padding: for layers where padding is active, the scheduler starts the central pixel at the top left corner and zero-pads the undefined edges of the activation window. In case of no padding, the scheduler starts the central pixel to the lower-right of the padded starting position. For all but the first layer in a network, the weight loading and computation phases overlap such that the weights for the next layer are pre-loaded to eliminate additional loading latency.

\begin{figure*}[t]
  \begin{center}
    \includegraphics[width=\linewidth]{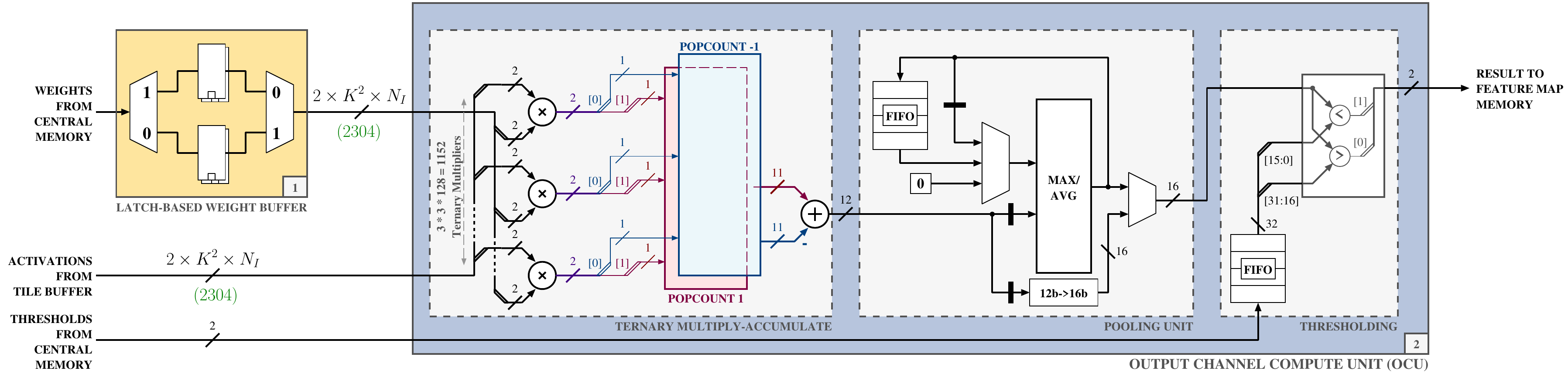}
    \caption{\revA{Block diagram of the compute units for the design point $K = 3$, $N_I = N_O = 128$, showing the dual inner weight buffers (1), used for double buffering to avoid load stalling, the OCU (2), including the completely unrolled multiply/add tree, computing 1\text{\textquotesingle}152 multiply-accumulate operations in a single cycle, the pooling block, which enables max and average pooling and the thresholding module used to ternarize intermediate results. Notably, the multiplier and popcounts are fully combinational and not pipelined, which adds to the energy efficiency of the compute core.}}
    \label{fig:ocu_pool}
  \end{center}
\end{figure*}

The \glspl{ocu} form the compute core of the accelerator. Figure \ref{fig:ocu_pool} shows the block diagram of a single \gls{ocu}.
Each \gls{ocu} contains two weight buffers, each of which is sized to hold all the kernel weights of one layer. Having two buffers allows executing the current layer while also loading the next layer's weights. The actual computations are done in the ternary multipliers, each of which computes one product of a single weight and activation. While the input trits are encoded in the standard two's complement format, the result of this computation is encoded differently, i.e. the encoding is given by $f$:
$$
f (x) = \left\{
\begin{array}{ll}
\revA{2'b10} & x = 1 \\
\revA{2'b01} & x = -1 \\
\revA{2'b00} & x = 0 \\
\end{array}
\right. 
$$

\begin{figure}
  \begin{center}
    \includegraphics[width=\linewidth]{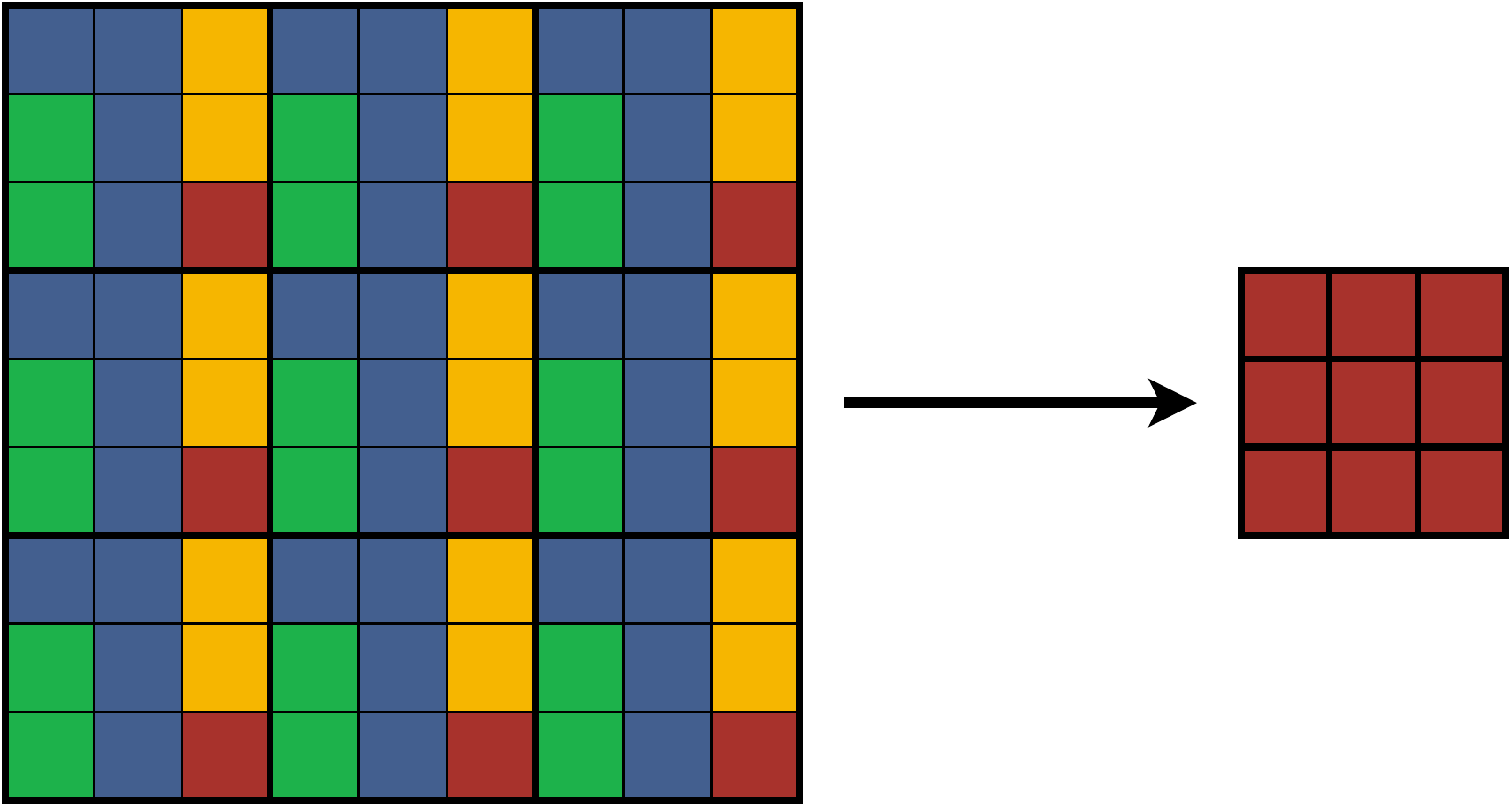}
    \caption{\revA{Example of pooling buffer scheduling for $9\times9$ feature maps applying $3\times3$ pooling. The feature map is traversed left-
    to-right, top-to-bottom. Blue pixels are stored in the pooling unit's register, yellow pixels are stored in the pooling unit's FIFO for later use and green pixels are loaded from the pooling unit's FIFO and compared to the current value. Best viewed in color.}}
    \label{fig:pooling}
  \end{center}
\end{figure}

This encoding allows calculating the sum of all multiplications by counting the number of ones in the MSB and subtracting the number of ones in the LSB of all results, which is done in the popcount modules. 
The resulting value is stored as an intermediate result, either for further processing with the pooling module or as input for the threshold decider. The threshold decider compares the intermediate values against two programmable thresholds and returns a ternary value, depending on the result of the comparison.
Notably, the \gls{ocu} is almost exclusively combinational, requiring only one cycle of latency for non-pooling layers. Registers are only used to silence the pooling unit and in the pooling unit itself to keep a running record of the current pooling window. Since every compute unit computes one output channel pixel at a time, there are no partial sums that have to be written back.\footnote{Which is a major difference from systolic arrays as well as output stationary designs!} However, to support pooling, each compute unit is equipped with a FIFO, a register, and an Add/Max ALU. In the case of max pooling, every newly computed value is compared to a previously computed maximum value for the window. In the case of average pooling, values are simply summed and the thresholds that are computed offline are scaled up accordingly. Figure \ref{fig:pooling} shows an example of the load \& store schedule for pooling operations.

Low-power optimizations have been made on all levels of the design, spanning from the algorithmic design of the neural networks over the system architecture down to the choice of memory cells.
\lstset{language=Python,style=myPython,caption={\revA{Loop unrolling of convolutional layers implemented in the CUTIE architecture. The highlighted lines 3-8 are computed in parallel in a single shot, in combinational logic.
    Each OCU computes one output pixel channel value, i.e. each OCU computes one instance of the third loop.}},label=lst:convcode, linebackgroundcolor={%
    \ifnum\value{lstnumber}>2
            \color{highlight}
    \fi
    }}
\begin{minipage}{\linewidth}
\begin{lstlisting}
for w in range(featuremap_width):
 for h in range(featuremap_height):
  for co in range(output_channels):
   for ci in range(input_channels):
    for kw in range(kernel_width):
     for kh in range(kernel_height):
      out_fm[w][h][co] += in_fm[w+kw][h+kh][ci]
      * kernel[kw][kh][ci][co]
\end{lstlisting}
\end{minipage}
Unlike most state-of-the-art architectures which use either systolic arrays or output-stationary scheduling approaches with iterative decomposition \cite{Chen2016, Conti2018, Andri2018, Chen2018, Jain2019, Sze2017, Moons2018, Knag2020}, the CUTIE architecture unrolls the compute architecture fully with respect to weight buffering and output pixel computation, such that no storing of partial results is necessary; each output channel value is computed in a single cycle, as shown in \revA{Listing \ref{lst:convcode}}.
The proposed design loads each data item exactly once and reduces overheads in multiplexing by clock gating unused modules. This applies to both the system level, with pipeline stages of the compute core that can be silenced, as well as to the module level, where the pooling module can be clock gated. To reduce both leakage and access energy, the feature map and weight memories can be implemented with standard cell latches, which are clock-gated down to the level of individual words. Generally, all flip-flops and latches in the design are clock-gated to reduce power consumption due to clock activity.



\subsection{Input Encoding}

To run real-world networks on the accelerator, the integer-valued input data has to be encoded with ternary values. We designed a novel ternary thermometer encoding based on the binary thermometer encoding \cite{Buckman2018}.
The binary thermometer encoding is an encoding function $f$, that maps an integer between 0 and M to a binary vector with M entries.
\begin{align*}
    f \colon \mathcal{N}_{M} &\to \mathcal{B}^{M} \\
    x &\mapsto f(x)
\end{align*}
\begin{equation*}
    f(x)_{i} = 
    \begin{cases}
        1 & i < x \\
        -1 & i \ge x
    \end{cases}
\end{equation*}

The ternary thermometer encoding is an encoding function $g$ that maps an integer between 0 and 2M to a ternary vector of size M.
\begin{align*}
    g \colon \mathcal{N}_{2M} &\to \mathcal{B}^{M} \\
    x &\mapsto g(x)
\end{align*}
$$g(x)_{i} = \mathrm{sgn}(x-M) \cdot\frac{ f(|x-M|)_{i} + 1 }{2}$$

The ternary thermometer encoding makes use of the additional value in the ternary number set with respect to the set of binary numbers and can encode inputs that are twice the size for a binary vector of a given size.
The introduction of 0s in the encoding scheme further helps to reduce toggling activity in the compute units, lowering the average energy cost per operation. 
As an example, for $M = 128$, and $x=110$ the binary thermometer encoding produces $\left[1\right]^{110}\left[-1\right]^{18}$, whereas the ternary thermometer encoding produces $\left[-1\right]^{18}\left[0\right]^{110}$.

\subsection{Exemplary Instantiations of CUTIE}\label{sec:instantiations}

The architecture of CUTIE is highly parametric. In the following, we present two practical embodiments of the general architecture, which we will then push to full implementation. The instantiations of the accelerator presented in this section can process convolutions with a kernel of size $3 \times 3$ or smaller, using a stride between (1,1) and (3,3) with independent striding for the width and height dimension. It further supports average pooling and maximum pooling. Both no padding and full zero-padding, i.e. padding value of size 1 on every edge of feature maps, are supported. 
Depending on the requirements of the application, the feature map memory size and weight memory size should be configured to store the largest expected feature map and network. For the sake of evaluating the architecture, we chose to implement one version that supports feature maps up to a size of $32 \times 32 $ pixels for both the current input feature map and the output feature map using \glspl{scm} and another version supporting sizes up to $160 \times 120$ feature map pixels using SRAMs. 
The supported feature map memory size does not restrict the functionality, since feature maps that do not fit within the memory can be processed in tiles. \revA{Assuming the feature maps need to be transfered from and to an external DRAM memory which requires \SI{20}{pJ/Bit}, several orders of magnitude more energy than accessing internal memory, the critical goal is to minimize the amount of data transfered from and to external memory. To achieve that, we propose to adopt the depth-first computing schedule described in \cite{Goetschalckx2019}. }

\revA{To estimate the energy cost of processing the feature map in tiles and to compare the layer-first and depth-first strategies on CUTIE, we compute the number of processed tiles per layer, the number of tiles that need to be transfered over the chip's I/O and the number of weight kernels that need to be switched for both the depth-first as well as the layer-first strategies. We assume a network consisting of eight convolutional layers using 3$\times$3 kernels and 128 input and output channels. Using these results and simulated energy costs for computations and memory transfers, we compute the additional cost when processing large feature maps layer- and depth-wise. For large frames, the cost is clearly dominated by the external memory access energy. Table \ref{tab:tiling} shows an exploration over different frame sizes starting from 32$\times$32 for which no tiling is required and extending to 64$\times$64 and 96$\times$96 that require significant external memory transfer. We find that by minimizing the feature map movement, the depth-first strategy consumes significantly less than the layer-first strategy for practical cases.}

While the CUTIE core is designed to be integrated with a host processor, one key idea to reduce system-level energy consumption realized in the architecture is the autonomous operation of the accelerator core. The control implementation allows the accelerator to compute a complete network without interaction with the host. In the presented version, the weight memories, the layer FIFO, and threshold FIFOs are designed to store up to eight full layers, which can be scheduled one after another without any further input. In general, the number of layers can be freely configured, at the cost of additional FIFO and weight memory.

Besides offering support for standard convolutional layers, the architecture can be used for depthwise convolutional layers by using weight kernels where each kernel is all zeros except for one channel. Further, it can be used for ternary dense layers with input size smaller or equal to $3 \times 3 \times 128 = 1\text{\textquotesingle}152$ and output size smaller or equal to 128 by mapping all dense layer matrix weights to the $3 \times 3 \times 128$ weight buffer of an \gls{ocu}.
\section{Implementation}\label{sec:implementation}
This section discusses the implementation of the CUTIE accelerator architecture. The results from physical layouts in a \SI{22}{\nano\meter} technology, one using \glspl{scm} and another using SRAMs, and from synthesis in a \SI{7}{\nano\meter} technology are presented and discussed.


\begin{table}
    \label{tab:tiling}
    \centering
    \caption{\revA{Estimated energy consumption of a network consisting of 8 convolutional layer without pooling for tiled computation of large feature maps on a GF 22 SCM implementation including I/O and external DRAM}}
    \begin{tabularx}{\linewidth}{X | r r}
      \textbf{} & \textbf{Depth-first} & \textbf{Layer-first} \\
      32$\times$32 & \SI{7.3}{\micro\joule} & \SI{7.3}{\micro\joule}\\
      \hspace{1.5mm}Bit accesses from or to external memory & \SI{209}{kB} & \SI{209}{kB} \\
      \hspace{1.5mm}Feature map transfer energy & \SI{4.2}{\micro\joule} & \SI{4.2}{\micro\joule} \\
      \hspace{1.5mm}Weight memory transfer energy & \SI{0.3}{\micro\joule} & \SI{0.3}{\micro\joule} \\
      \hspace{1.5mm}Computational energy & \SI{2.8}{\micro\joule} & \SI{2.8}{\micro\joule} \\
      \\
      64$\times$64 & \SI{277}{\micro\joule} & \SI{1069}{\micro\joule}\\
      \hspace{1.5mm}Bits moved from or to external memory & \SI{12.6}{MB} & \SI{52.8}{MB} \\

      \hspace{1.5mm}Feature map transfer energy & \SI{252}{\micro\joule} &\SI{1057}{\micro\joule} \\
      \hspace{1.5mm}Weight memory transfer energy & \SI{2.5}{\micro\joule} & \SI{0.3}{\micro\joule} \\
      \hspace{1.5mm}Computational energy & \SI{22.5}{\micro\joule} & \SI{11.5}{\micro\joule} \\
      \\
      96$\times$96 & \SI{3734.5}{\micro\joule} &  \SI{6030.3}{\micro\joule}\\
      \hspace{1.5mm}Bit accesses from or to external memory & \SI{179.3}{MB} & \SI{300.1}{MB} \\
      \hspace{1.5mm}Feature map transfer energy & \SI{3586}{\micro\joule} & \SI{6002}{\micro\joule} \\
      \hspace{1.5mm}Weight memory transfer energy & \SI{14.5}{\micro\joule} & \SI{0.3}{\micro\joule} \\
      \hspace{1.5mm}Computational energy & \SI{134}{\micro\joule} & \SI{28}{\micro\joule} \\
    \end{tabularx}
    \label{tab:my_label}
\end{table}

\subsection{Interface Design}

The interface of the accelerator consists of a layer instruction queue and read/write interfaces to the feature map and weight memories. The interface is designed to allow integration into a \gls{soc} design targeting near-sensor processing. In this context, a pre-processing module could be connected to a sensor interface, with a host processor only managing the initial setup and off-chip communication. This setup consists of writing the weights into their respective weight memories and pre-loading the layer instructions into the instruction queue. In the actual execution phase, i.e. once data is loaded continuously, the accelerator is designed to autonomously execute the layer instructions without needing any further input besides the input feature maps and return only a highly-compressed feature map or even final labels. The end of computation is signalled by a single-bit interrupt to the host.

\subsection{Dimensioning}

\begin{figure}
    \begin{center}
        \includegraphics[width=\linewidth, trim=5 30 0 0]{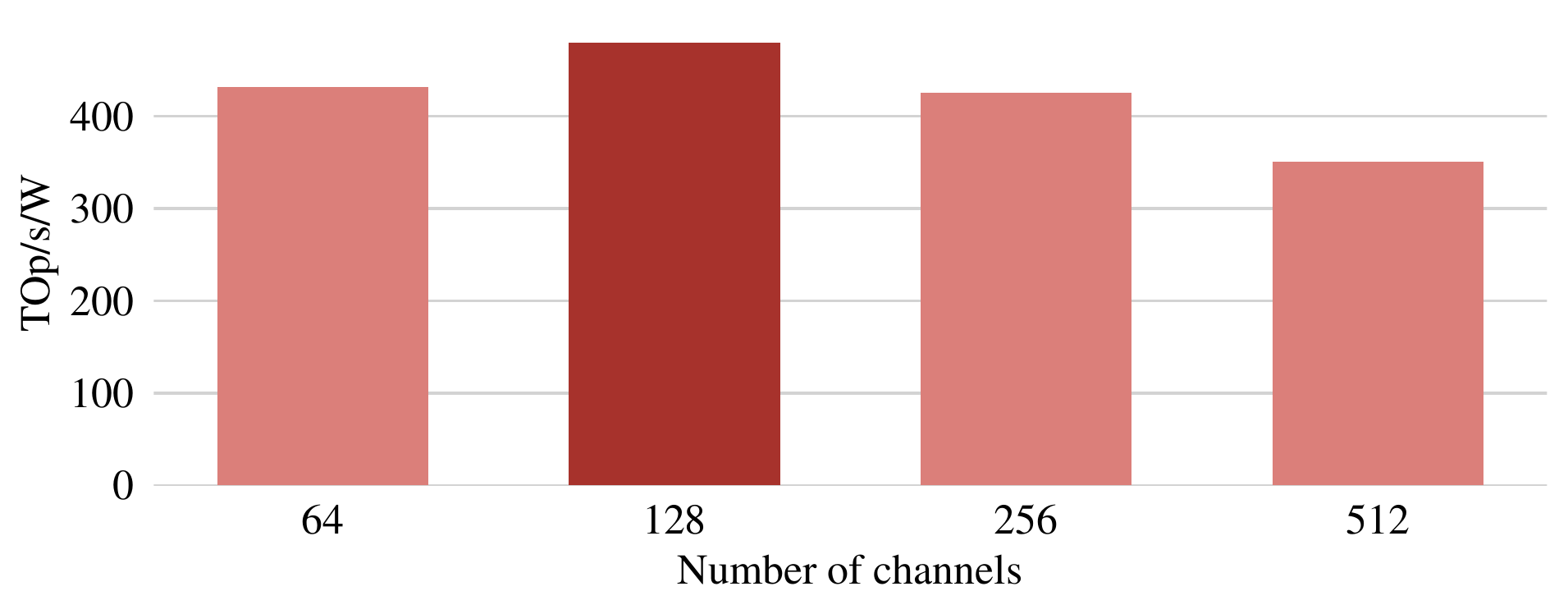}
    \end{center}
    \caption{\revA{Estimation of accelerator-level energy efficiency using data from the simulation of single OCUs, assuming SCM-based memories. Feature maps and weights were drawn from a uniform random distributions. There is a peak in energy efficiency at 128 channels before falling off for increasing channel numbers.}}
    \label{fig:channelsvsefficiency}
\end{figure}

The CUTIE architecture is not architecturally constrained to support a certain number of input/output channels, i.e. it can be parameterized to support an arbitrary amount of channels.
Since it can be synthesized with support for any number of channels and feature map sizes, the proposed implementation was designed to optimize the accuracy vs. energy efficiency trade-off for the CIFAR-10 dataset. To this end, the compute units were synthesized and routed for different channel numbers to evaluate the impact of channel number on the energy efficiency of individual compute units and by extension, the whole accelerator. 
The estimations were performed for 64, 128, 256, and 512 channels. 
To estimate the energy efficiency of the individual implementations, a post-layout power simulation was performed, using randomly generated activations and weights. This experiment was repeated and averaged over 300 cycles, i.e. 300 independently randomly generated weight tensors and feature maps were used. Further, post-synthesis simulation estimations for the energy cost of memory accesses, encoding \& decoding, and the buffering of activations and weights were added. 
The estimations for the resulting accelerator-level energy efficiency are shown in Figure \ref{fig:channelsvsefficiency}. Since these estimations were made using a post-layout power simulation of a single \gls{ocu}, they take into account the wiring overheads introduced by following the completely unrolled compute architecture. One of the main drivers for lower efficiency in the designs with more channels is the decrease in layout density and an increase in wiring overheads.
While energy efficiency per operation does not directly imply energy per inference, it is a strong indicator of system-level efficiency.

\subsection{Implementation Metrics}

\begin{figure}
    \begin{center}
        \includegraphics[width=\linewidth, trim=50 30 50 12]{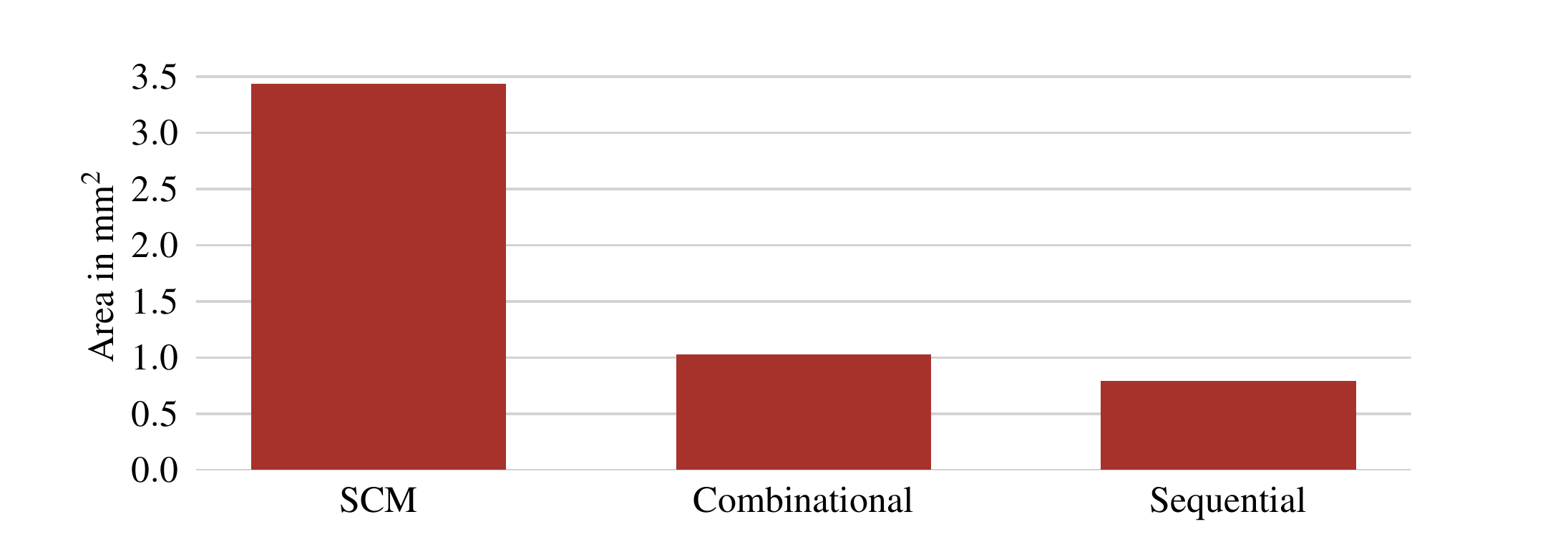}
    \end{center}
    \caption{\revA{Breakdown of the area usage of the SCM implementation of the accelerator core in \SI{22}{\nano\meter} technology. The majority of the area is used by the standard cell memories, which are used to store feature maps and weight kernels. Clock area is \revA{negligibly} small, due to deliberate low clock speeds and hierarchical clock gating}}
    \label{fig:areabeakdown}
\end{figure}

The accelerator design was implemented with a full backend flow in GlobalFoundries \SI{22}{\nano\meter} FDX and synthesized in TSMC \SI{7}{\nano\meter} technology. 
The first of two implementations based on GlobalFoundries \SI{22}{\nano\meter} FDX was synthesized using SRAMs supplied with \SI{0.8}{\volt} for feature map and weight memories and 8 track standard cells operating at \SI{0.65}{\volt}. The second of the GF \SI{22}{\nano\meter} implementations uses \gls{scm}-based feature map and weight memories as well as 8 track standard cells for its logic cells, all supplied with \SI{0.65}{\volt}.
The TSMC \SI{7}{\nano\meter} implementation similarly uses \gls{scm}-based memories to allow for voltage scaling.
The post-synthesis timing reports show that the GF \SI{22}{\nano\meter} implementations should be able to operate at up to \SI{250}{MHz}. We chose to run both the SCM as well as the SRAM implementation at a very conservative frequency of \SI{66}{MHz}. Since we did not run a full backend implementation of the \SI{7}{\nano\meter} version, we chose to estimate the performance at the same clock frequency and voltage as the \SI{22}{\nano\meter} versions.
The total area required by the design is \SI{7.5}{mm^2} for both \SI{22}{\nano\meter} implementations and approximately \SI{1.2}{mm^2} at a layout density of 0.75 for the \SI{7}{\nano\meter} implementation. The reason for both GF \SI{22}{\nano\meter} implementations requiring the same amount of area is due to the larger memories supported in the SRAM implementation, as explained in section \ref{sec:instantiations}. A breakdown of the area usage in the \gls{scm}-based \SI{22}{\nano\meter} implementation is shown in Figure \ref{fig:areabeakdown}.

For the GF \SI{22}{\nano\meter} implementations, the sequential and memory cells take up around 80\% of the overall design's area, while the clock buffers and inverters constitute only a very small amount of the total area. This characteristic is due to the choice of using latch-based buffers for a lot of the design and clocking the accelerator at a comparatively low frequency, while also extensively making use of clock-gating at every level of the design's hierarchy.
Note that even though the area of the design is storage-dominated, power and energy are not, which is one of the key reasons for the extreme energy efficiency of CUTIE.

\section{Results and Discussion}\label{sec:results}
\label{ch:results}

This section discusses the evaluation results of the proposed accelerator design. First, we discuss the design and training of the network that is used to evaluate the accelerator's performance. Next, we discuss the general evaluation setup. Finally, we present the implementation and performance metrics and compare our design to previous work.

\subsection{Quantized Network Training}

The accelerator was evaluated using a binarized and a ternarized version of a neural network, using the binary thermometer encoding and the ternary thermometer encoding for input encoding.
The network architecture is shown in Table \ref{tab:networkoverview}.

\begin{table}
    \centering
    \caption[Layer architecture of the tested CNN]{Layer architecture of the tested CNN}
    \renewcommand*\arraystretch{1}
    \begin{tabularx}{\linewidth}{X | r r c c}
    \textbf{Layer} & \textbf{Input Dim} & \textbf{Op} & \textbf{Kernel} & \textbf{Padding} \\
    \hline
        2D Convolution & 126$\times$32$\times$32 & \SI{297}{MOp} & 3$\times$3 & (1,1) \\
        2D Convolution & 128$\times$32$\times$32 & \SI{302}{MOp} & 3$\times$3 & (1,1) \\
        2D Convolution & 128$\times$32$\times$32 & \SI{302}{MOp} & 3$\times$3 & (1,1) \\
        Max Pooling & 128$\times$32$\times$32 & - & 2$\times$2 & \revA{(0,0)} \\
        2D Convolution & 128$\times$16$\times$16 & \SI{75.5}{MOp} & 3$\times$3    & (1,1) \\
        2D Convolution & 128$\times$16$\times$16 & \SI{75.5}{MOp} & 3$\times$3 & (1,1) \\
        Max Pooling & 128$\times$16$\times$16 & - & 2$\times$2 & \revA{(0,0)} \\
        2D Convolution & 128$\times$8$\times$8 & \SI{18.9}{MOp} & 3$\times$3 & (1,1) \\
        2D Convolution & 128$\times$8$\times$8 & \SI{18.9}{MOp} & 3$\times$3 & (1,1) \\
        Max Pooling & 128$\times$8$\times$8 & - & 2$\times$2 & \revA{(0,0)} \\
        2D Convolution & 128$\times$4$\times$4 & \SI{4.7}{MOp} & 3$\times$3 & (1,1) \\
        Avg Pooling & 128$\times$4$\times$4 & - & 4$\times$4 & \revA{(0,0)} \\
        Fully connected & 128 & 2.6 KOp & - & - \\
        \textbf{Total} & - & \textbf{1.1 GOp} & - & - \\
    \end{tabularx}
    \label{tab:networkoverview}
\end{table}


Each convolutional layer is followed by a batch normalization layer and a Hardtanh activation \cite{Collobert2011} layer.
For the quantized versions of the network, the activation layer is followed by a ternarization layer. The preceding convolutional layer, batch normalization layer and Hardtanh activation layer are merged into a single Fused Convolution layer. Any succeeding pooling layers are then merged as well. 
The reason for using Hardtanh activations over, for example, the more popular ReLU activation which is also usually used in \glspl{bnn} is the inclusion of all three ternary values in the range of the function. We further found that the Hardtanh activation converged much more reliably than the ReLU activation for the experiments we ran.
\revA{We have tested networks with depthwise-separable convolutions in place of standard convolutions but have found that accuracy decreases substantially when ternarizing these networks, which is in line with the results in \cite{Phan2020} and \cite{Phan2020_2}. Further, depthwise-separable convolutions require twice the feature map data movement, while performing fewer operations overall. Since CUTIE's architecture greatly reduces the cost of the elementary multiply and add operations, the cost of accessing local buffers is relatively high. Hence, layers that have been optimized in a traditional setting to minimize the number of operations are not guaranteed to be energy efficient.}



The approach for training the networks taken in this work is based on the INQ algorithm \cite{Zhou2017}. Training is done in full-precision for a certain number of epochs, after which a pre-defined ratio of all weights are quantized according to a quantization schedule. These two steps are iterated until all weights are quantized. One degree of freedom in this algorithm is the order in which the weights are quantized, called the quantization strategy. 
 We evaluated three quantization strategies for their impact on accuracy, and sparsity, which is linked to energy efficiency for execution on the proposed architecture. 
The strategies evaluated in this work are the following:

\begin{itemize}
    \item Magnitude: Weights are sorted in descending order by their absolute value
    \item \revA{Magnitude-Inverse}: Weights are sorted in ascending order by their absolute value
    \item \revA{Zig-Zag}: Weights are sorted by taking the remaining smallest and largest values one after another. 
\end{itemize}

\begin{figure}
 \begin{center}
      \includegraphics[width=\linewidth, trim= 0 30 0 0]{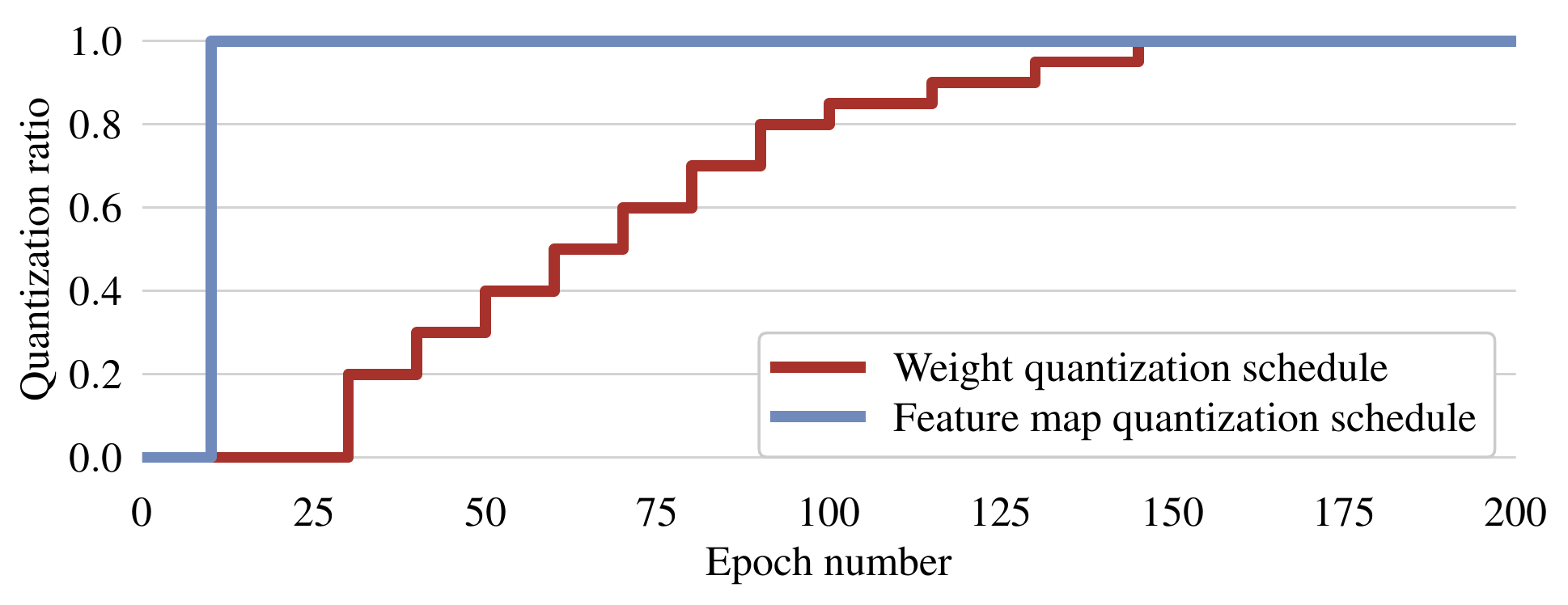}
    \end{center}
    \caption{\revA{Quantization schedule for the presented network. Weights and feature map pixels are quantized separately, using different schedules. The weight quantization schedule uses a decaying step size, which starts at 20\%, decreases to 10\% and finishes with 5\% of all weights. }}
    \label{fig:quantizationschedule}
\end{figure}

For both the ternarized and binarized versions, the weights were quantized using the quantization schedule shown in Figure \ref{fig:quantizationschedule}. 
The CIFAR-10 dataset was used for training and the CIFAR-10 test data set was used for all evaluations.
The network was trained using the ADAM optimizer \cite{Kingma2014} over a total of 200 epochs. 

\subsection{Evaluation Setup}

In addition to the quantized network, a testbench was implemented to simulate the cycle-accurate behavior of the accelerator core. The testbench generates all necessary signals to load all weights and feature maps into the accelerator core and load the layer instructions into the layer FIFO.
The \SI{22}{\nano\meter} implementations were simulated using annotated switching activities from their respective post-layout netlist to simulate the average power consumption of the accelerator core, including memories, during the execution of each layer. Analogously, the \SI{7}{\nano\meter} implementation was simulated using its post-synthesis netlist.
For power simulation purposes, each layer was run separately from the rest of the network. This guarantees that each loading phase is associated with its layer, which is required to properly estimate the energy consumption of a layer.  
For throughput and efficiency calculations, the following formula for the number of operations in convolutional layers is used:

$$ \Gamma = 2 \cdot I_{W} \cdot I_{H} \cdot K \cdot K \cdot N_{I} \cdot N_{O}  $$

where K corresponds to the side length of the convolutional kernel, \revA{$I_W$ and $I_H$} are the output features maps' width and height, and $N_{I}$ \& $N_{O}$ are the input and output channel number, respectively.
$\Gamma$ corresponds to the number of additions and multiplications required to compute each output pixel, i.e. operations for pooling and activations are not considered.
Furthermore, the runtime of each layer is measured between the loading of the layer instruction and the write operation for the last output feature map pixel.

\subsection{Experimental Results}\label{sec:experimentalresults}

\begin{figure*}[t!]
    \begin{minipage}{\linewidth}
      \begin{center}
        \includegraphics[width=\linewidth, trim=120 0 120 0]{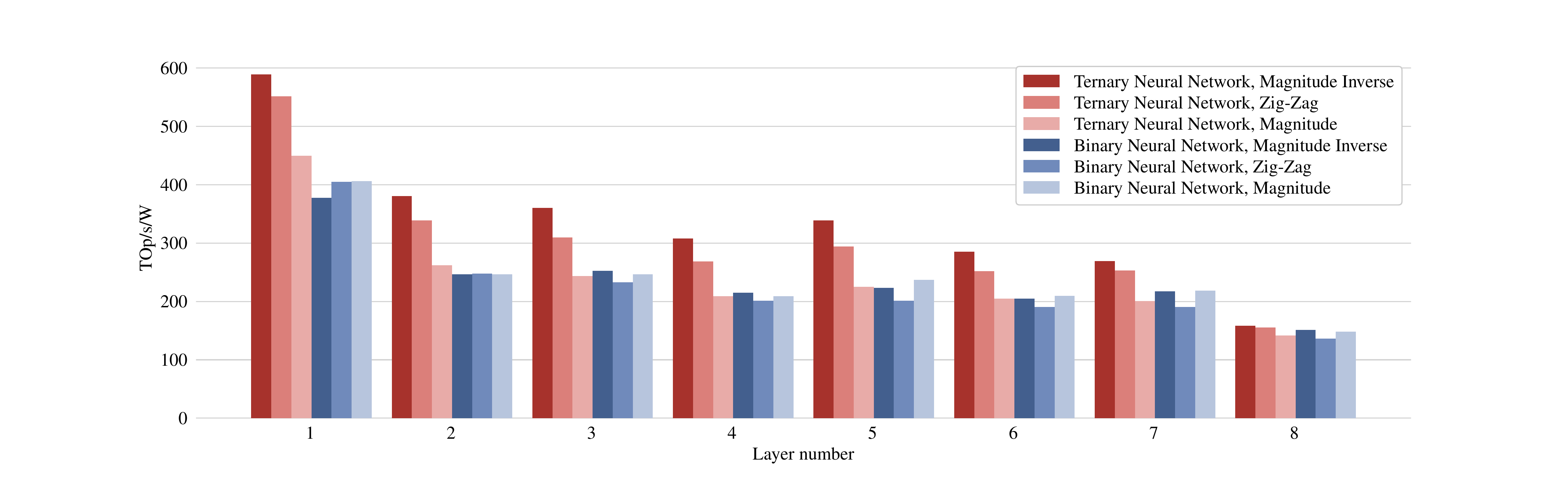}
      \end{center}
    \end{minipage}%
    \caption{\revA{Energy efficiency simulation results on the CIFAR-10 test dataset for the binarized \& ternarized networks comparing the different quantization strategies using the GF \SI{22}{\nano\meter} post-layout power simulation data. Notably, the energy efficiency per operation increases with increasing sparsity of the weight kernels as shown in table \ref{tab:ternarystrategyoverview}.}}
    \label{fig:simulation} 
\end{figure*}

The energy per operation for the \SI{22}{\nano\meter} implementation using different quantization strategies is shown in Figure \ref{fig:simulation}. The energy efficiency scales almost linearly with the sparsity of the executed network. 
This trend can be explained by zeros in the adder trees leading to nodes not toggling, which results in lower overall activity.

\begin{figure}
    \begin{center}
      \includegraphics[width=\linewidth, trim=0 20 0 0]{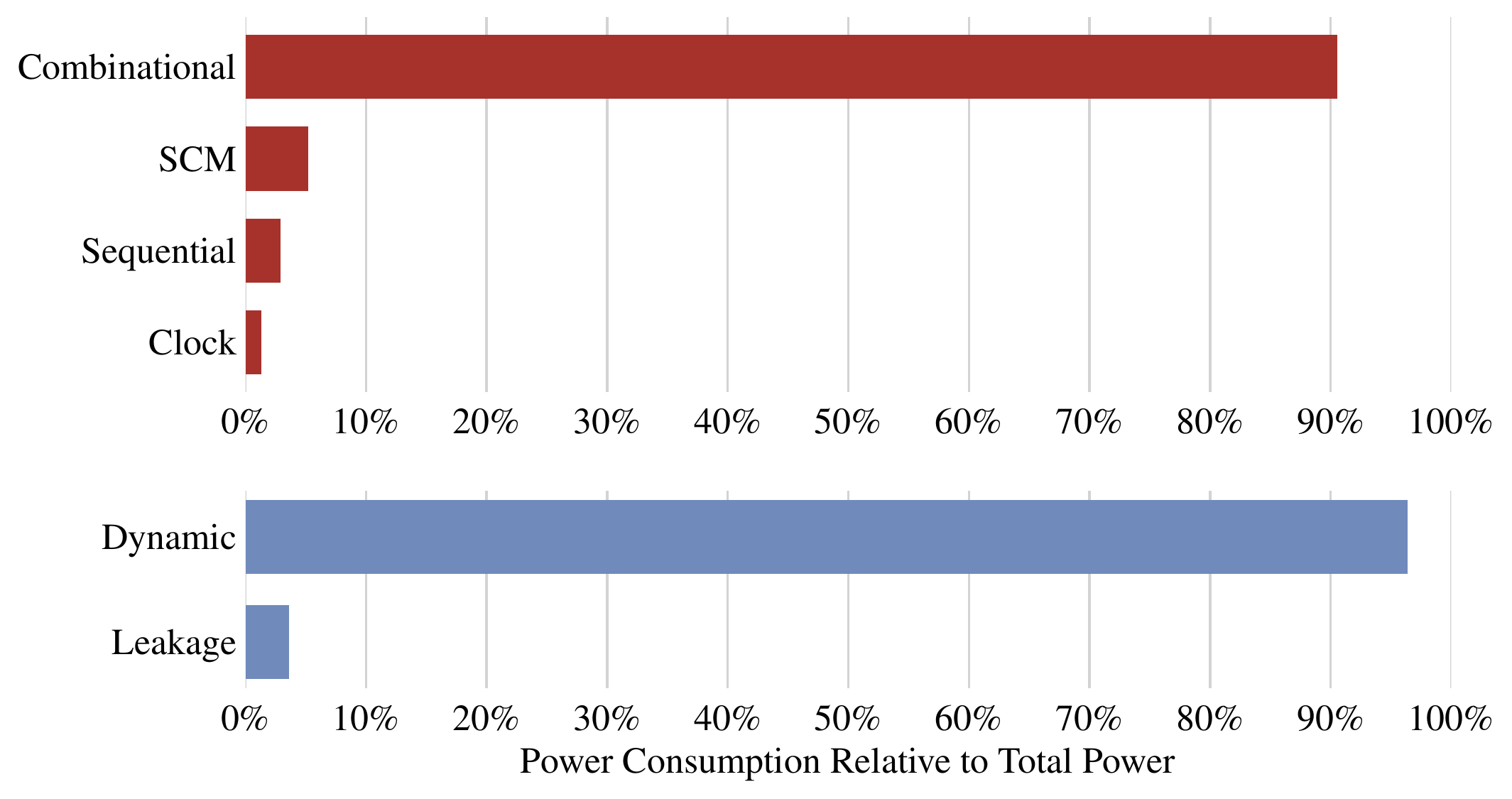}
    \end{center}
    \caption{\revA{Power breakdown of the accelerator core implementation in \SI{22}{\nano\meter} technology with SCM-based feature map and weight memories, running the Magnitude-Inverse trained ternary network. The overall power is clearly dominated by combinational cells, where over 90\% of the total power is spent.}} 
    \label{fig:powerbreakdown}
\end{figure}

A breakdown of power consumption by cell type, as well as by dynamic and leakage power is shown in Figure \ref{fig:powerbreakdown}. The static power consumption makes up 4.6\% of the overall power consumption in the \SI{22}{\nano\meter} implementation, most of which stems from the SCMs. 
Notably, the power consumption is dominated by combinational cells which underlines the effectiveness of the architecture, since this implies most energy is spent in computations, rather than memory accesses or transfers. 

\begin{figure}[t!]
    \begin{minipage}{\linewidth}
        \includegraphics[width=\linewidth]{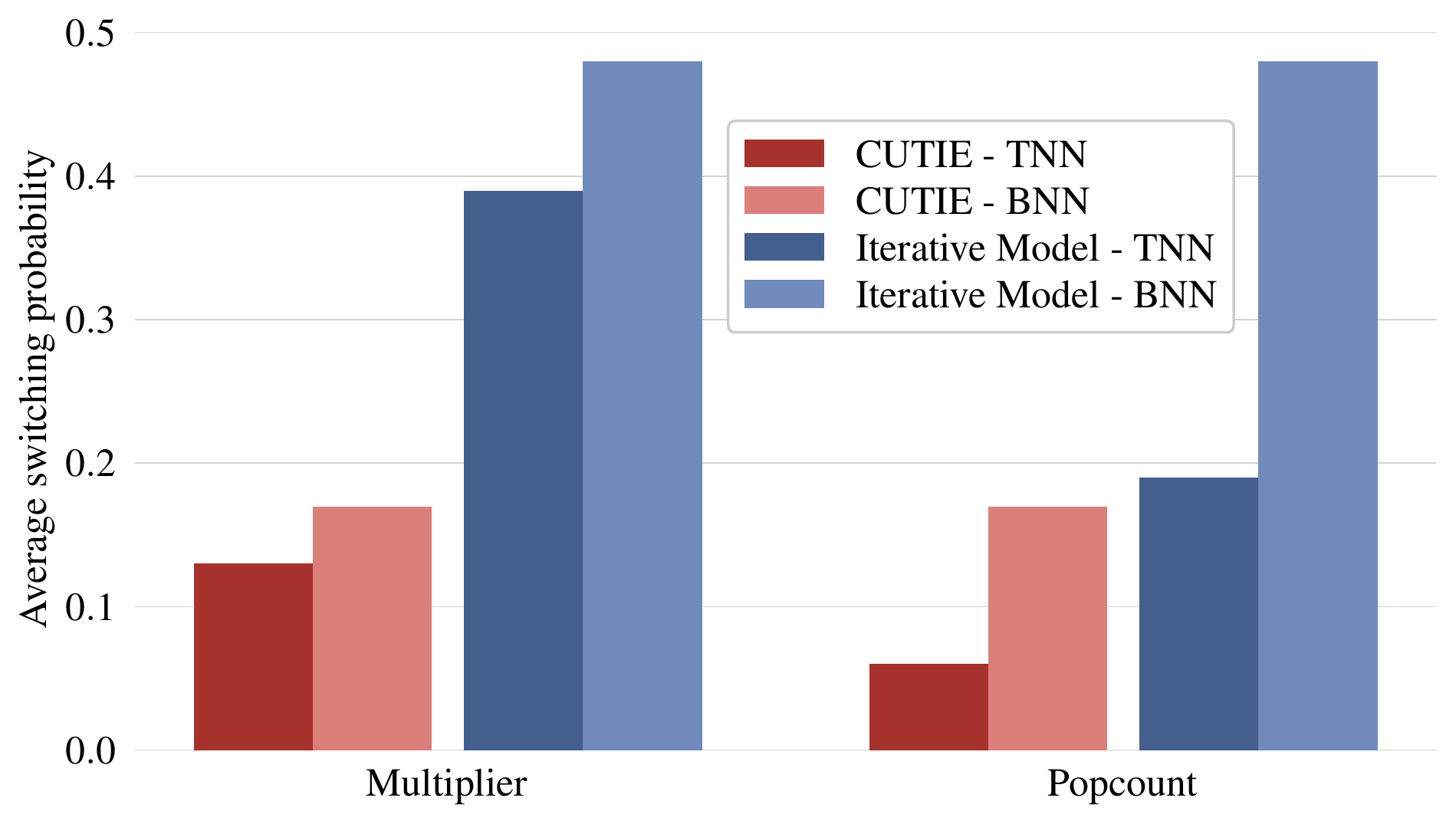}
    
    \end{minipage}%
    
    \caption{\revA{Overview of the switching probabilities at the multiplier and adder tree input nodes respectively, smaller is better.  For the binary case, toggling in the multipliers directly translates to switching activity in the adder trees, while for the ternary case the sparsity of the network reduces switching activity at the adder tree input nodes by $\approx2\times$. Moreover, the smoothness of feature maps is exploited by unrolling the compute units, which is reflected in a $\approx3\times$ smaller switching probability compared to an iteratively decomposed model. Best viewed in color.}}
    \label{fig:smoothnessstuff}
\end{figure}

The analysis of the per-layer energy efficiency for both binary and ternary neural networks reveals a sharp peak in the first layer, which can be explained with the structural properties of the thermometer encoding, i.e. the first feature map contains 66.3\% zeros on average.  Furthermore, with the decreasing number of operations in deeper layers, the energy cost of loading the weights increase in proportion to the energy cost of computations, which explains the decreasing energy efficiency in deeper layers. 

The binary thermometer encoding and ternary thermometer encoding were compared for their use with the ternarized network version. The results show that the ternary thermometer encoding provides a small increase between 0.5\% and 1.5\% in test accuracy, while energy efficiency is kept within 2\% of the binary thermometer. Further, the drop in accuracy between the \revA{32-bit} full-precision version and the ternary version can be reduced to as little as 3\%.

Finally, the ternary network trained with the \revA{Magnitude-Inverse} quantization strategy using the ternary thermometer encoding was evaluated on the post-synthesis netlist of the \SI{7}{nm} implementation, achieving a peak energy efficiency of \SI{3140}{TOp/s/W} in the first layer and an average efficiency of \SI{2100}{TOp/s/W}.

\subsection{Comparison of Quantization Strategies}\label{sec:comparisonquantization}

An overview of test accuracy and sparsity for all tested strategies is given for the binarized and ternarized versions in Table \ref{tab:ternarystrategyoverview}.

\begin{table}
    \centering
    \caption{Impact of quantization strategy on test accuracy and sparsity for binarized \& ternarized networks on the CIFAR-10 dataset evaluated in the \SI{22}{\nano\meter} SCM implementation}
    \renewcommand*\arraystretch{1.1}
    \begin{tabularx}{\linewidth}{X | r r r}
         & Accuracy & Weighty Sparsity & Avg. TOp/s/W \\
         \hline
         \textbf{Full-Precision} & 91\% & - & - \\
         \textbf{Ternary, TT$^{*}$} \\
         Magnitude & 86.5\% & 7.4\% & \SI{260}{TOp/s/W}\\
         Magnitude-Inverse & 87.4\% & 60.7\% & \SI{392}{TOp/s/W} \\
         Zig-Zag & 88.1\% & 49.1\% & \SI{345}{TOp/s/W}\\
         \textbf{Ternary, BT$^{*}$} \\   
         Magnitude & 85.9\% & 6.9\% & \SI{262}{TOp/s/W}\\
         Magnitude-Inverse & 86.8\% & 60.8\% & \SI{399}{TOp/s/W} \\
         Zig-Zag & 86.6\% & 49.2\% & \SI{342}{TOp/s/W}\\
         \textbf{Binary} \\
         Magnitude & 83.3\% & 0\% & \SI{240}{TOp/s/W}\\
         Magnitude-Inverse & 80.1\% & 0\% & \SI{248}{TOp/s/W}\\
         Zig-Zag & 82.8\% & 0\% & \SI{229}{TOp/s/W}\\
    \end{tabularx}
    \smallskip
    \parbox[t]{\textwidth}{\footnotesize
    
      $^{*}$ BT: Binary Thermometer \\
      $^{*}$ TT: Ternary Thermometer
    }
    \label{tab:ternarystrategyoverview}
\end{table}

The energy per inference for the most efficient ternary version in \SI{22}{\nano\meter} adds up to \SI{2.8}{\micro \joule}, the energy per inference for the best binary version to about \SI{4.4}{\micro \joule}.
These results allow three observations: first, the quantization strategy not only impacts the accuracy of the resulting network but also the distribution of weights - the number of zeros for the \revA{Magnitude-Inverse} strategy is more than 8x higher than for Magnitude, at comparable accuracy.
The second observation is that energy efficiency increases significantly for very sparse networks. The \revA{Magnitude-Inverse} strategy trains a network that runs 36\% more efficiently than the one trained with Magnitude for the ternary case.
Lastly, the results imply that the optimal quantization strategy might be different for the binary and ternary case. 
Most importantly, for all training experiments we have run, we have found that ternary neural networks consistently outperform their binary counterparts on the CUTIE architecture by a considerable margin, both in terms of accuracy, with 5\% higher test accuracy, as well as in terms of energy efficiency, with 36\% lower energy per inference.

\subsection{Exploiting Feature Map Smoothness}
By fully unrolling the compute units with respect to the feature map channels and weights, we reduce switching activity in the adder tree of the compute units by an average of 66.6\% with respect to architectures that use an output-stationary approach and iterative decomposition. 
Iteratively decomposed architectures require the accelerator to compute partial results on partial feature maps and weight kernels. The typical approach to implement this is tiling the feature map and weight kernels in the input channel direction, and switch the weight and feature map tiles every cycle. This leads to much higher switching activity. 

In the ternary case, an input node of the adder tree switches when the corresponding weight value is non-zero and the feature map value changes. Calculating the mean number of value switches between neighboring pixels, we found that the binary feature map pixels have an average Hamming distance of 44 out of 256 bit and the ternary feature map pixels have an average pixel-to-pixel Hamming distance of 33 out of 256 bit following the 3-ary encoding of CUTIE. 
It exploits this fact by keeping the weights fixed for the execution of a full layer, which eliminates switching activity due to changing the weight tile while a previous feature map tile is scheduled. To quantify this effect, we analyzed the switching activity of the presented network trained with all quantization strategies on an output-stationary iterative architecture model, taking into account the network weights as well. Figure~\ref{fig:smoothnessstuff} shows the occurring switching activity for CUTIE versus a model with $2\times$ iterative decomposition for the binary \revA{Magnitude} and ternary \revA{Magnitude-Inverse} trained networks.

\subsection{Comparison of Binary and Ternary Neural Networks}

Since the set of ternary values includes the set of binary values, a superficial comparison between binary and ternary neural networks on the proposed accelerator architecture is fairly straight-forward, as binary neural networks can be run on the accelerator as-is. To fairly compare, however, it is important to discount certain contributions that only appear because the accelerator core supports ternary operations.
Most importantly, the overhead in memory storage, accesses, encoding, and decoding should be subtracted, as well as the energy spent in the second popcount module.
To apply these considerations on the architecture, the following simplifications are made:

\begin{itemize}
    \item The power used for memory accesses is divided by 1.6.
    \item The power used in the popcounts of the compute units is halved.
    \item The power used for encoding and decoding is subtracted.
\end{itemize}

While these reductions do not account for all differences between the ternary and a binary implementation of the accelerator, they give a reasonably close estimate, considering that the power spent in popcounts, memories and encoding \& decoding modules accounts for around 80\% of the total power budget. Adding up the reductions, an average of around 30\% should be subtracted from the measured values of the GF \SI{22}{\nano \meter} \gls{scm} implementation to get an estimate for the energy efficiency of a purely binary version of the accelerator. Even including this discount factor into all calculations, the energy of the binary neural network would be reduced to around \SI{3}{\micro J}, which is slightly higher than the ternary version. 
Taking into account that the achieved accuracy for the ternary neural network comes in at around 88\% while the binary version achieves around 83\%, the ternary implementation is both more energy-efficient and more accurate in terms of test accuracy than the binary version.

\begin{table*}[b]
\caption{Comparison of the proposed architecture to state-of-the-art accelerators}  
  \begin{tabularx}{\textwidth}{X|r|r|r|r|r|r r r}
    & \cite{Andri2020} & \cite{Moons2018} & \cite{Bankman2019} &  \cite{Knag2020} &  \cite{Jain2019} & \multicolumn{3}{c}{\textbf{This work}}  \\
    \hline
    Computation Method & digital & digital & mixed  & digital & analog & digital & digital & digital \\
    Weight Precision & binary & binary & binary  & binary & ternary & ternary & ternary & ternary \\
    Activation Precision & binary & binary & binary & binary & ternary & ternary & ternary & ternary\\
    Memory Implementation & SCM & SRAM & SRAM & SCM & SRAM & SRAM & SCM & SCM\\
    Technology & \SI{22}{\nano\meter} & \SI{28}{\nano\meter} & \SI{28}{\nano\meter}  & \SI{10}{\nano\meter} & \SI{32}{\nano\meter} & \SI{22}{\nano\meter} & \SI{22}{\nano\meter} & \SI{7}{\nano\meter} \\
    Core Area [$\text{mm}^2$] & 0.7 & 1.4 & 5.76  & \textbf{0.39} & 1.96 & 7.5 & 7.5 & $1.2^{b}$ \\
    Core Voltage [V]& 0.4 & 0.66 & 0.6  & \textbf{0.37} & - & 0.65 & 0.65 & 0.65\\
    Peak Throughput [TOp/s] &  0.3 & 2.8 & -  & \textbf{160} & 114 & 16 & 16 & 16 \\
    Peak Core Energy Efficiency [TOp/s/W] & 223 & 230 & - & 617 & - & 457 & 589 & \textbf{3'140} \\
    Average Core Energy Efficiency [TOp/s/W] & 36 & 145 & 772 & 617 & 127 & 305 & 392 & \textbf{2'100} \\
    Accuracy on CIFAR-10 & 87\% & 86\% & 85.6\%  & $86\%^{a}$ & - & \textbf{88\%} & \textbf{88\%} & \textbf{88\%} \\
    Energy per Inference on CIFAR-10 [\SI{}{\micro J}] (excl. I/O) & 1.3--7.3 & 13.86 & 2.61 & 3.2 & - & 3.6 & 2.8  & \textbf{0.52} \\
  \end{tabularx}
  \label{tab:comparison}
  \smallskip
  \parbox[t]{\textwidth}{\footnotesize
    $^a$: uses same network as \cite{Moons2018} 
    $^b$: expected value at 0.75 cell layout density
    }
\end{table*}

\subsection{Comparison with the State-of-the-Art}\label{sec:comparisonsoa}

A comparison of our design with similar accelerators cores is shown in Table \ref{tab:comparison}. The implementation in TSMC \SI{7}{\nano\meter} technology outperforms even the most efficient digital binary accelerator design, implemented in comparable Intel \SI{10}{\nano\meter} technology as reported by Knag et al. \cite{Knag2020}, by a factor of at least 3.4$\times$ in terms of energy efficiency per operation and 5.9$\times$ in terms of energy per inference as well as the most efficient mixed-signal design as reported by Bankman et al. \cite{Bankman2019}, requiring a factor of 4.8$\times$ less energy per inference. 

For a fairer comparison to other state-of-the-art accelerators, we also report post-layout simulation results in GF \SI{22}{\nano\meter} technology, which similarly outperforms comparable implementations as reported in Moons et al. \cite{Moons2018} by a factor 2.5$\times$, both in terms of peak efficiency as well as average efficiency per operation. The more practical comparison between the energy per inference on the same data set reveals that our design outperforms all other designs by an even larger margin, i.e. by at least 4.8$\times$, while even increasing the inference accuracy with respect to all other designs. 
However, our design is less efficient in terms of throughput per area compared to other state-of-the-art designs. This is a deliberate design choice, which is due to the unrolled architecture of CUTIE.

\section{Conclusion}\label{sec:conclusion}

In this work, we have presented three key ideas to increase the core efficiency of ultra-low bit-width neural network accelerators and evaluated their impact in terms of energy per operation by combining them in an accelerator architecture called CUTIE. 
The key ideas are:
1) completely unrolling the data path with respect to all feature map and filter dimensions to reduce data transfer cost and switching activity by making use of spatial feature map smoothness, 
2) moving the focus from binary neural networks to ternary neural networks to capitalize on the inherent sparsity and
3) tuning training methods to increase sparsity in neural networks at iso-accuracy.
Their combined effect boosts the core efficiency of digital binary and ternary accelerator architectures and contribute to what is to the best of our knowledge the first digital accelerator to surpass POp/s/W energy efficiency for neural network inference. 

Future work will focus on extending the core architecture to enable efficient computation of different layers and integrating the accelerator core into a sensor system-on-chip.

\section*{Acknowledgement}

The authors would like to thank \textit{armasuisse Science \& Technology} for funding this research. This project was supported in part by the EU's H2020 Programme under grant no. 732631 (OPRECOMP).

\bibliography{./main}
\vfill\eject

\begin{IEEEbiography}[{\includegraphics[width=1in,height=1.25in,clip,keepaspectratio]{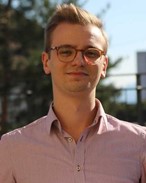}}]{Moritz Scherer} received the B.Sc. and M.Sc. degree in electrical engineering and information technology from ETH Zürich in 2018 and 2020, respectively, where he is currently pursuing a Ph.D. degree at the Integrated Systems Laboratory. His current research interests include the design of ultra-low power and energy-efficient circuits and accelerators as well as system-level and embedded design for machine learning and edge computing applications.
Moritz Scherer received the ETH Medal for his Master’s thesis in 2020.
\end{IEEEbiography}

\begin{IEEEbiography}[{\includegraphics[width=1in,height=1.25in,clip,keepaspectratio]{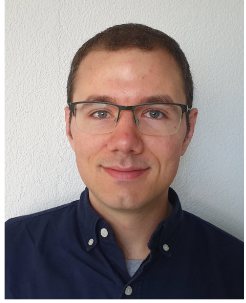}}]{Georg Rutishauser} received his B.Sc. and M.Sc.degrees in Electrical Engineering and Information Technology from ETH Zürich in 2015 and 2018,respectively. He is currently pursuing a Ph.D. degree at the Integrated Systems Laboratory at ETH Zürich. His research interests include algorithms and hardware for reduced-precision deep learning, and their application in computer vision and embedded systems.
\end{IEEEbiography}

\begin{IEEEbiography}[{\includegraphics[width=1in,height=1.25in,clip,keepaspectratio]{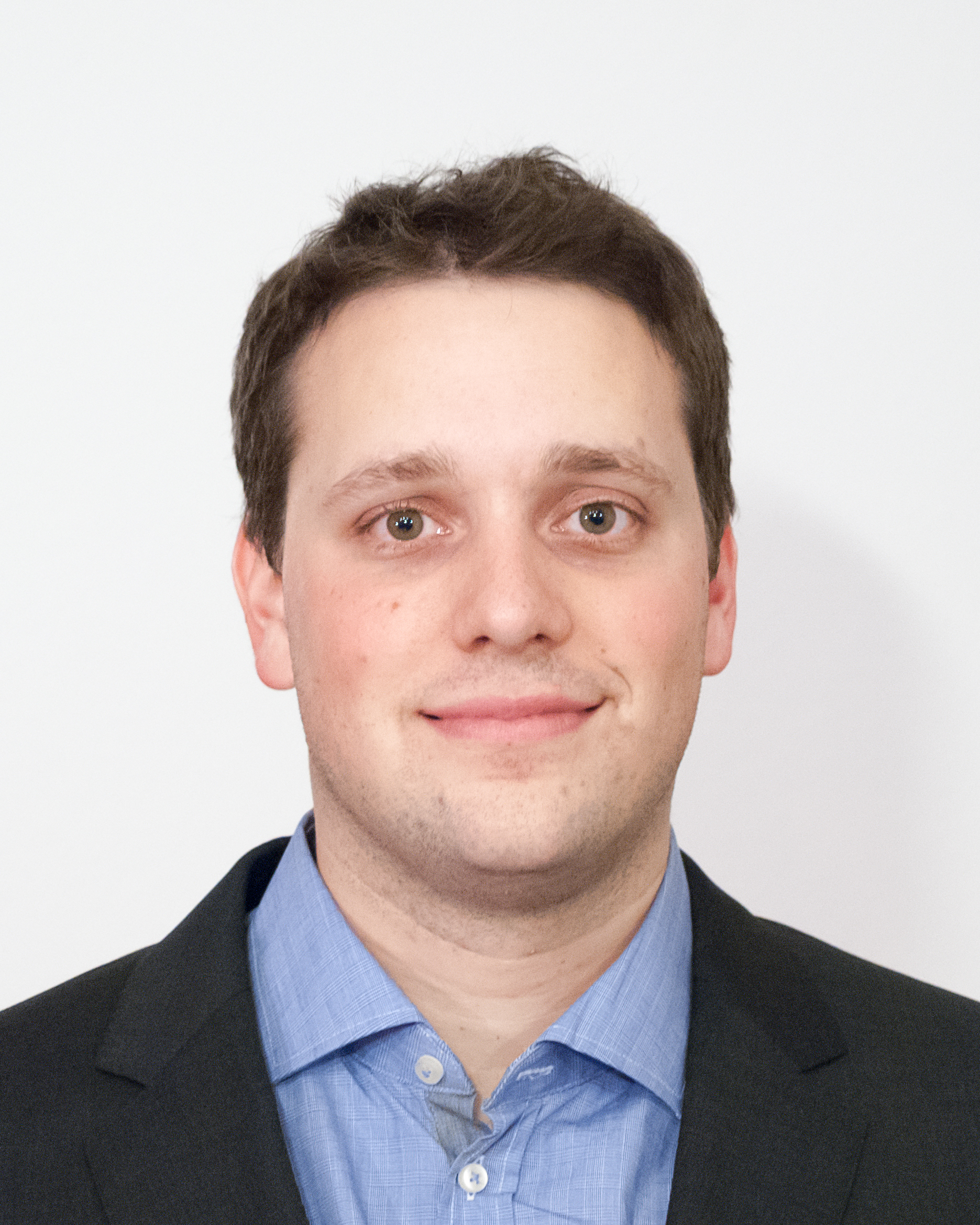}}]{Lukas Cavigelli}
received the B.Sc., M.Sc., and Ph.D. degree in electrical engineering and information technology from ETH Zürich, Zürich, Switzerland in 2012, 2014 and 2019, respectively. After spending another year as a Postdoc at ETH Zürich, he has joined Huawei's Zurich Research Center in Spring 2020. His research interests include deep learning, computer vision, embedded systems, and low-power integrated circuit design. He has received the best paper award at the VLSI-SoC and the ICDSC conferences in 2013 and 2017, the best student paper award at the Security+Defense conference in 2016, the ETH Medal for his Ph.D. thesis in 2019, and the Donald O. Pederson best paper award (IEEE TCAD) in 2019.
\end{IEEEbiography}

\begin{IEEEbiography}[{\includegraphics[width=1in,height=1.25in,clip,keepaspectratio]{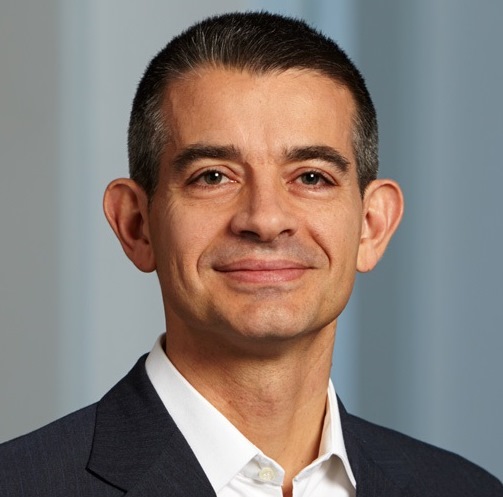}}]{Luca Benini}
is the Chair of Digital Circuits and Systems at ETH Zürich and a Full Professor at the University of Bologna. He has served as Chief Architect for the Platform2012 in STMicroelectronics, Grenoble. Dr. Benini’s research interests are in energy-efficient system and multi-core SoC design. He is also active in the area of energy-efficient smart sensors and sensor networks. He has published more than 1’000 papers in peer-reviewed international journals and conferences, four books and several book chapters. He is a Fellow of the ACM and of the IEEE and a member of the Academia Europaea.
\end{IEEEbiography}

\end{document}